\documentclass[twocolumn,aps,prb,superscriptaddress,nofootinbib]{revtex4-1}
\usepackage{graphicx}
\usepackage{times}
\usepackage{bm}
\usepackage[linktocpage=true]{hyperref}
\usepackage{pgfplots}
\usepackage{amsmath}
\usepackage{braket}
\usetikzlibrary{arrows}
\usepackage{enumerate}
\usetikzlibrary[patterns] 
\usetikzlibrary[snakes] 
\usetikzlibrary[shapes] 
\usepackage{pgfplots}

\begin{document}

\title{Tunneling conductance in semiconductor-superconductor hybrid structures}

\author{John Stenger}
\affiliation{Department of Physics and Astronomy, West Virginia University, Morgantown, WV 26506}
\author{Tudor D. Stanescu}
\affiliation{Department of Physics and Astronomy, West Virginia University, Morgantown, WV 26506}
\affiliation{Condensed Matter Theory Center and Joint Quantum Institute, Department of Physics, University of Maryland, College Park, Maryland, 20742-4111, USA}

\begin{abstract}
We study the differential conductance for charge tunneling into a semiconductor wire--superconductor hybrid structure, which is actively investigated as a possible scheme for realizing topological superconductivity and Majorana zero modes. The calculations are done based on a tight-binding model of the heterostructure using both a  Blonder-Tinkham-Klapwijk approach and a Keldysh non-equilibrium Green's function method. The dependence of various tunneling conductance features on the coupling strength between the semiconductor and the superconductor, the tunnel barrier height, and temperature is systematically investigated. We find that treating the parent superconductor as an active component of the system, rather than a passive source of Cooper pairs, has qualitative consequences  regarding the low-energy behavior of the differential conductance. In particular, the presence of sub-gap states in the parent superconductor, due to disorder and finite magnetic fields, leads to characteristic particle-hole asymmetric features and to the breakdown of the quantization of the zero-bias peak associated with the presence of Majorana zero modes localized at the ends of the wire. The implications of these findings for the effort toward the realization of Majorana bound states with true non-Abelian properties are discussed.
\end{abstract}

\maketitle

\section{Introduction}

The search for quantum states of matter characterized by nontrivial topological properties has gained significant momentum in recent years motivated, in part, by the discovery of topological insulators \cite{Hasan2010,Qi2011,Bernevig2013} and by the exciting prospect of  using topological objects, such as the Majorana zero modes, \cite{Alicea2012,Leijnse2012,Beenakker2013,Stanescu2013} as a platform for fault tolerant quantum computation. \cite{Nayak2008,Stanescu2017} The proposal for engineering topological superconductors capable of supporting Majorana zero modes that has generated the most promising experimental results involves a semiconductor nanowire with strong spin-orbit coupling proximity-coupled to a standard \textit{s}-wave superconductor in the presence of an applied magnetic field.\cite{Lutchyn2010,Oreg2010} Upon increasing the magnetic field, the system undergoes a topological quantum phase transition\cite{Read2000,Sau2010}  from a topologically-trivial superconductor to a topological superconductor with mid-gap Majorana zero modes localized near the ends of the wire. The presence of the Majorana bound states can be probed by tunneling charge into the end of the wire, which is predicted to generate a zero-bias conductance peak that is quantized ($2e^2/h$) at zero temperature.\cite{Sengupta2001,Law2009,Flensberg2010,Wimmer2011,Fidkowski2012} These theoretical proposals and predictions have led to an impressive experimental effort\cite{Mourik2012,Deng2012,Das2012,Finck2013,Chang2015,Albrecht2016,Zhang2016,Deng2016,Chen2016} for realizing Majorana zero modes in semiconductor-superconductor hybrid structures. The results of this effort are promising, but a conclusive demonstration of the realization of Majorana bound states in the laboratory remains elusive. 

A serious problem is the persistence of significant discrepancies between theory and experiment.  While the theoretical possibility of realizing topological superconductivity and Majorana zero modes in semiconductor-superconductor structures that satisfy certain requirements is beyond any reasonable doubt, understanding how well real devices realized in the laboratory satisfy these requirements  remains a serious challenge. In addition to careful systematic measurements, overcoming this challenge calls for theoretical studies that go beyond idealized conditions and incorporate all relevant details associated with real devices. Since charge tunneling is, so far, the preferred method of detecting the presence of Majorana zero modes, significant theoretical effort was devoted to numerical and analytical studies of the tunneling differential conductance using both normal metal\cite{Stanescu2011,Pientka2012,Prada2012,Lin2012,Rainis2013,Stanescu2014,Yan2014,Setiwan2015,Liu2017}  
and superconductor\cite{Peng2015,Sharma2016,Chevallier2016,Setiawan2017} leads. 
Most of these studies, however, do not consider explicitly the parent s-wave superconductor; instead, its role is reduced to that of a passive source of Cooper pairs for the semiconductor nanowire. Recently, it has been emphasized that the parent superconductor  should be treated  as an active ingredient in the physics of the heterostructure.\cite{Reeg2017,Stanescu2017a}  This treatment of the semiconductor wire and the parent superconductor on equal footing leads to a renormalization of the low-energy modes in the hybrid system\cite{Stanescu2017a} and the emergence of additional conductance peaks at energies corresponding to the bulk gap of the parent superconductor.\cite{Reeg2017}

In this work we investigate theoretically the tunneling conductance of a normal-metal -- semiconductor-wire -- superconductor hybrid structure and demonstrate that the explicit treatment of the parent superconductor as an active component of the system has major consequences not only at  energies on the order of the bulk superconducting gap, but all the way down to zero energy. Our findings are based on a systematic numerical analysis of a tight-binding model of the heterostructure. The conductance calculations are carried out  within  both the  Blonder-Tinkham-Klapwijk (BTK) theory\cite{Blonder1982} and the Keldysh non-equilibrium Green's function formalism\cite{Rammer1986}. To gain a better physical understanding of the low-energy features revealed by the differential conductance, we  compare these features with the local density of states. We establish the qualitative features that characterize the  dependence of tunneling conductance on  the coupling strength between the semiconductor wire and the parent superconductor, the strength of the tunnel barrier, and temperature. We also consider the possibility of sub-gap states being present in the parent superconductor and identify the consequences of these states hybridizing with low-energy states in the wire. We find that the presence of sub-gap states leads to characteristic particle-hole asymmetric features in the low-energy differential conductance and to a breakdown of the zero-bias peak quantization at zero temperature. Our findings  suggest that a detailed modeling of the parent superconductor, including the possible presence of disorder-induced sub-gap states in the presence of a finite magnetic field, is a strong requirement for any attempt to quantitatively (and, at this stage, even qualitatively) reproduce the experimentally  measured  differential conductance.

The remainder of this paper is organized as follows. In Sec. \ref{S_II} we introduce a tight-binding model for the normal-metal -- semiconductor-wire -- superconductor hybrid structure (Sec. \ref{S_IIA}) and briefly describe the BTK-type formalism (Sec. \ref{S_IIB}) and the Keldysh approach (Sec. \ref{S_IIC}) for calculating the differential conductance. The results of our numerical analysis are presented in Sec. \ref{S_III}. First, we 
compare the two approaches for calculating the tunneling conductance and establish the correspondence between the differential conductance (in the low tunneling regime) and the local density of states (Sec. \ref{S_IIIA}). The dependence of the low-energy features on the transparency of the semiconductor-superconductor interface and the tunnel barrier height is discussed in Sec. \ref{S_IIIB}. We show that additional features associated with charge transport via quasiparticles in the parent superconductor occur at energies larger than the bulk superconducting gap. The effects of disorder-induced sub-gap states in the parent superconductor are investigated in Sec.\ref{S_IIIC} and the dependence of the Majorana-induced zero-bias conductance peak on  the relevant parameters (i.e., tunnel barrier height, temperature, and sub-gap states density) is summarized in Sec. \ref{S_IIID}. Our conclusions are presented in Sec. \ref{S_IV}.

\section{Theoretical model} \label{S_II}

In this section we introduce the tight-binding model for the  normal-metal -- semiconductor-wire -- superconductor hybrid structure and briefly describe the BTK and Keldysh methods for calculating the differential conductance. Both approaches explicitly incorporate the parent superconductor.

\subsection{Tight-binding Hamiltonian}  \label{S_IIA}
We consider charge tunneling from a normal metal lead into a semiconductor wire - superconductor hybrid structure through a controllable potential barrier.  The Hamiltonian describing the system has the following generic form:
\begin{equation}
H = H_m+ H_{sm} + H_{sc} + T_m+T_{sc} + H_{ext},                            \label{Htot}
\end{equation}
where $ H_m$,  $H_{sm}$, and  $H_{sc}$ describe the normal metal, semiconductor wire, and superconductor components, respectively, the next two terms describe the coupling between the semiconductor wire and the metallic lead ($T_m$) and the coupling between the wire and the parent superconductor ($T_{sc}$), while $H_{ext}$ includes contributions from external fields, such as magnetic fields and electrostatic potentials.
The specific form of each term in Eq. (\ref{Htot}) depends on on the degree of complexity that we want to incorporate into the model. In this study we consider a simple tight-binding description of the hybrid system involving a lattice consisting of coupled parallel chains. Specifically, we have
\begin{equation}
H_m = \sum_{i,\delta}t_{m}^{\delta}{\rm a}^{\dagger}_{i}{\rm a}_{i+\delta}+ \mu_m\sum_{i}{\rm a}^{\dagger}_{i}{\rm a}_{i},
\end{equation}
where $i=(i_x,i_y)$ labels the position, with $1\leq i_y\leq N_y$ designating the chain and $1\leq i_x \leq N_m$ the position along the chain, $\delta=(\pm 1, 0)$ or $(0,\pm 1)$ is a nearest neighbor vector, $t_m^{\delta}=(t_{m}^x,t_{m}^y)$ is the hopping matrix element, and $\mu_m$ the chemical potential of the metallic lead. Using spinor notation, the electron creation operator on site $i$ is ${\rm a}_i^\dagger =({\rm a}_{i\uparrow}^\dagger  ~{\rm a}_{i\downarrow}^\dagger)$. The semiconductor wire (including the applied external fields) is described by the tight-binding Hamiltonian
\begin{eqnarray}
H_{sm}+H_{ext} &=& \sum_{i, \delta}t_{sm}^{\delta}c^{\dagger}_{i}c_{i+\delta}+\sum_{i}(\mu_{sm}+V_i)c^{\dagger}_{i}c_{i} \nonumber \\
&+& i\sum_{i}[\frac{\alpha_x}{2}c^{\dagger}_{i+\delta_x}\hat{\sigma}_y c_i -\frac{\alpha_y}{2}c^{\dagger}_{i+\delta_y}\hat{\sigma}_x c_i+h.c.] \nonumber \\
&+& \Gamma\sum_{i}c^{\dagger}_{i}\hat{\sigma}_x c_{i},
\end{eqnarray}
where $c_i^\dagger=(c_{i\uparrow}^\dagger  ~c_{i\downarrow}^\dagger)$ is the electron creation operator on the site $i=(i_x, i_y)$ of the semiconductor wire, with $1 \leq i_x \leq N_{sm}$ and  $1\leq i_y\leq N_y$, 
$t_{sm}^{\delta}=(t_{sm}^x,t_{sm}^y)$ is the nearest-neighbor hopping,
$\delta=(\pm 1, 0)$ or $(0,\pm 1)$ is a nearest neighbor vector, while $\delta_x=(1, 0)$ and $\delta_y=(0, 1)$ are (positive) unit vectors in the \textit{x}- and \textit{y}-direction, respectively,
 $\mu_{sm}$ is the chemical potential, and $(\alpha_x,\alpha_y)$ are the Rashba spin-orbit coupling coefficients. The position-dependent potential $V_i$ describes the tunneling barrier and additional back-gate-induced potentials, while $\Gamma$ represents the Zeeman splitting. The matrices $\hat{\sigma}_\mu$, with $\mu = x,y,z$,  represent Pauli matrices associated with the spin degree of freedom.
The parent superconductor is described within the Bogoliubov-de Gennes (BdG) formalism by
\begin{eqnarray}
H_{sc} &=& \sum_{i,\delta}t_{sc}^{\delta}a^{\dagger}_{i}a_{i+\delta} + \mu_{sc}\sum_{i}a^{\dagger}_{i}a_{i}  \nonumber \\
&+& \Delta_{0} \sum_{i}(a_{i\uparrow}^{\dagger}a^{\dagger}_{i\downarrow}+a_{i\downarrow}a_{i\uparrow}),   
\end{eqnarray}
where $a_i^\dagger=(a_{i\uparrow}^\dagger  ~a_{i\downarrow}^\dagger)$ is the electron creation operator on the site  $i=(i_x, i_y)$ of the superconducting chains, with $1 \leq i_x \leq N_{sc}$ and  $1\leq i_y\leq N_y$,
$t_{sc}^{\delta}=(t_{sc}^x,t_{sc}^y)$ is the nearest-neighbor hopping matrix element, $\mu_{sc}$ the chemical potential of the superconductor, and $\Delta_0$ is the superconducting gap. We assume that the superconductor is equal to or longer than the semiconductor wire, $N_{sc} \geq N_{sm}$.
The coupling between the semiconductor wire and the metallic lead is described by the term
\begin{equation}
\label{msm}
T_m= \tilde{t}_{m-sm}\sum_{i_y=1}^{N_y}\left({\rm a}_{(N_m, i_y)}^\dagger c_{(1,i_y)} + h.c.\right), 
\end{equation}
where the nearest-neighbor hopping $\tilde{t}_{m-sm}$ quantifies the coupling strength, while $(N_m, i_y)$ and $(1,i_y)$ label the sites at the right end of the metallic lead and the left end of the semiconductor wire, respectively.
Finally, the coupling between the semiconductor and the superconductor is described by the term
\begin{equation}
T_{sc}=\tilde{t}\sum_{i_x=1}^{N_{sm}}\sum_{i_y=1}^{N_y}\left(c_{i}^\dagger a_{i} + a_{i}^\dagger c_{i}\right), 
\label{msm}
\end{equation}
where $i=(i_x,i_y)$, $\tilde{t}$ is the hopping across the semiconductor-superconductor (SM-SC) interface, while $c_i$ and $a_i$ designate annihilation operators on the SM and SC sides, respectively. 

It may be convenient to use a Green's function formalism and calculate the effective Green function for the semiconductor wire by integrating out the degrees of freedom of the parent superconductor.\cite{Stanescu2013} The proximity effect induced by the superconductor is captured by  a self-energy term,
\begin{equation}
\Sigma_{sc}(\omega)=\tilde{t}^2G_{sc}(\omega),  \label{Sig_sc}
\end{equation}
where $G_{sc}(\omega)$ is the Green function of the parent superconductor at the SM-SC interface. 
Assuming that the parent superconductor is a truly bulk system (i.e., wide enough and thick enough), the self-energy becomes local,\cite{Stanescu2013} $\Sigma_{sc}(\omega; i, j) = \delta_{i,j}\Sigma_{sc}(\omega)$, with
\begin{equation}
\Sigma_{sc}(\omega)=-|\tilde{t}|^2\nu_{sc}\left(\frac{\omega\hat{\tau}_0+\Delta_0\hat{\tau}_x}{\sqrt{\Delta_0^2-\omega^2}}+\zeta\hat{\tau}_z\right), \label{fullscse}
\end{equation}
where $\nu_{sc} = \sqrt{4t_{sc}\mu_{sc}-\mu_{sc}^2}/({2 t_{sc}^2})$ is the surface density of states of the bulk superconductor at the chemical potential and $\zeta= (2t_{sc}\mu_{sc}-4t_{sc}^2)/(\mu_{sc}^2-4t_{sc}\mu_{sc})$ is a proximity-induced shift of the SM chemical potential. The matrices $\hat{\tau}_\mu$, with $\mu = x,y,z$,  represent Pauli matrices associated with the particle-hole degree of freedom.
We define the effective SM-SC coupling as 
\begin{equation}
\gamma = |\tilde{t}|^2\nu_{sc}. \label{gamma}
\end{equation}
Note that here we do not discuss the physics associated with the finite wire thickness in the direction perpendicular to the SM-SC interface (i.e., the $z$-direction). This is critical when considering electrostatic effects, such as those generated by back-gate potentials. In general, the effective SM-SC coupling $\gamma$ is a band-dependent quantity  proportional to the amplitude squared of the wave-functions (corresponding to a given confinement-induced band) at the SM-SC interface and may involve band off-diagonal contributions.\cite{Stanescu2013} In the weak-coupling limit, $\gamma/\Delta_0\rightarrow 0$, the self-energy (\ref{fullscse}) can be approximated at low energy by the anomalous contribution
\begin{equation}
\label{Din}
\Sigma_{sc}= -\gamma~\!\tau_x.
\end{equation}
Note that in this approximation the proximity-induced gap (defined as the minimum quasiparticle gap in the absence of an applied magnetic field) is $\Delta_{ind} = \gamma$ and the minimum critical Zeeman field associated with the topological quantum phase transition is $\Gamma_c=\gamma\equiv \Delta_{ind}$. However, in general the critical field and the induced gap have different values.\cite{Sticlet2016,Stanescu2017a} 
To describe the low-energy spectral properties of the hybrid structure it is convenient to use the density of states (DOS) $\rho(\omega)$ and the local density of states (LDOS) $\bar{\rho}(i,\omega)$ at the end of the wire. In terms of the effective wire Green's function $G(\omega) = [\omega - H_{sm} -\Sigma_{sc}(\omega)]^{-1}$ we have  
\begin{eqnarray}
\bar{\rho}(i, \omega) &=& -\frac{1}{\pi}{\rm Im}[G(i,\omega)], \\
\rho(\omega) &=& \sum_{i}\bar{\rho}(i, \omega).
\end{eqnarray}

\subsection{BTK Formalism} \label{S_IIB}

The calculation of the differential conductance using the  Blonder-Tinkham-Klawijk (BTK) formalism\cite{Blonder1982} involves  calculating the reflection and transmission coefficients for incoming (and outgoing) plane waves by solving the Bogoliubov-de Gennes (BdG) equation for the total Hamiltonian given by Eq. (\ref{Htot}) with appropriate boundary conditions. The Schr\"odinger equation that determines the reflection and transmission coefficients is 
\begin{eqnarray}
\label{shro}
&~&\sum_{j_x=0}^{N_{tot}+1}\sum_{j_y=1}^{N_y}\sum_{\sigma^\prime}({\cal H}_{i\sigma,j\sigma^\prime}-\omega\delta_{i,j}\delta_{\sigma,\sigma^\prime})\Psi_{j\sigma^\prime}=0 \label{SchEq} \\
&~&{\rm for}~ ~ i_x=1,\dots,N_{tot}, ~~i_y=1,\dots, N_y, ~~\sigma=\pm 1, \nonumber
\end{eqnarray}
where ${\cal H}$ is the first quantized BdG Hamiltonian corresponding to Eq. (\ref{Htot}) and $N_{tot} = N_m+N_{sm}+N_{sc}$. For convenience, we define the transverse modes $\phi_\nu$, with $1\leq \nu\leq N_y$, characterized by the wave functions $\phi_\nu(i_y)=\sqrt{2/(N_y+1)}\sin[i_y \nu \pi/(N_y+1)]$. Each transport channel is labeled by the pair of quantum numbers $(\nu~\!\sigma)$ corresponding to a given transverse mode and spin orientation and the reflection and transmission coefficients are $2N_y\times 2N_y$ matrices with matrix elements indexed by these channel labels. 

Consider now an incoming plane wave in channel $(\nu,\sigma)$. To ensure that the wave function entering the normal lead is a (propagating) plane wave, we impose (propagating) boundary conditions involving the first two sites of each chain, i.e., $j_x=0$ and $j_x=1$. Note that similar propagating boundary conditions involving two sites at the rightmost end of the superconductor, i.e., $j_x=N_{tot}$ and $j_x=N_{tot}+1$, correspond to an outgoing plane wave.  Specifically, the (incoming)  boundary conditions at the left end of the metallic lead are 
\begin{eqnarray}
\Psi_{(0,j_y)}^{(\nu,\sigma)} &=& \phi_\nu(j_y)\left(
\begin{array}{c} \delta_{\sigma,\uparrow} \\ \delta_{\sigma, \downarrow} \\ 0 \\ 0 \end{array} \right) +\sum_{\nu^\prime} \phi_{\nu^\prime}(j_y)\left(
\begin{array}{c} {[r_N]_{\nu\sigma, \nu^\prime\uparrow}} \\  {[r_N]_{\nu\sigma, \nu^\prime\downarrow}} \\  {[r_A]_{\nu\sigma, \nu^\prime\uparrow}} \\  {[r_A]_{\nu\sigma, \nu^\prime\downarrow}} \end{array} \right),  \nonumber \\
\Psi_{(1,j_y)}^{(\nu,\sigma)} &=& \phi_\nu(j_y)\left(
\begin{array}{c} \delta_{\sigma,\uparrow} \\ \delta_{\sigma, \downarrow} \\ 0 \\ 0 \end{array} \right) e^{i k_e^\nu a}  \label{rcoeff} \\
&~&~~~~~~~~~~~~+ \sum_{\nu^\prime} \phi_{\nu^\prime}(j_y)\left(
\begin{array}{c} {[r_N]_{\nu\sigma, \nu^\prime\uparrow}~ e^{-i k_e^{\nu^\prime} a}} \\  {[r_N]_{\nu\sigma, \nu^\prime\downarrow}~ e^{-i k_e^{\nu^\prime} a}} \\  {[r_A]_{\nu\sigma, \nu^\prime\uparrow}~ e^{i k_h^{\nu^\prime} a}} \\  {[r_A]_{\nu\sigma, \nu^\prime\downarrow}~ e^{i k_h^{\nu^\prime} a}} \end{array} \right),  \nonumber
\end{eqnarray}
where $r_N$ and $r_A$ are the normal and anomalous reflection coefficients, respectively, $a$ is the lattice constant, while $k_e^\nu$ and $k_h^\nu$ are wave vectors in the normal lead at energies $\omega$ and $-\omega$,  respectively,
\begin{equation}
k_{e(h)}^\nu(\omega) = \cos^{-1}\left(-\frac{\mu_m+\epsilon_\nu\pm\omega}{2t_m}\right),
\end{equation}
with $\epsilon_\nu = 2t_m\cos[\nu \pi/(N_y+1)]$. Similarly, the outgoing plane wave  dictates the boundary conditions at the rightmost end of the bulk superconductor chains, 
\begin{eqnarray}
\Psi_{(N_{tot}+1,j_y)}^{(\nu,\sigma)} &=& \sum_{\nu^\prime} \phi_{\nu^\prime}(j_y)\left(
\begin{array}{c} {[t_N]_{\nu\sigma, \nu^\prime\uparrow}} \\  {[t_N]_{\nu\sigma, \nu^\prime\downarrow}} \\  {[t_A]_{\nu\sigma, \nu^\prime\uparrow}} \\  {[t_A]_{\nu\sigma, \nu^\prime\downarrow}} \end{array} \right), \label{tcoeff} \\
\Psi_{(N_{tot},j_y)}^{(\nu,\sigma)} &=& \sum_{\nu^\prime} \phi_{\nu^\prime}(j_y)\left(
\begin{array}{c} {[t_N]_{\nu\sigma, \nu^\prime\uparrow}~ e^{-i q_e^{\nu^\prime} a}} \\  {[t_N]_{\nu\sigma, \nu^\prime\downarrow}~ e^{-i q_e^{\nu^\prime} a}} \\  {[t_A]_{\nu\sigma, \nu^\prime\uparrow}~ e^{i q_h^{\nu^\prime} a}} \\  {[t_A]_{\nu\sigma, \nu^\prime\downarrow}~ e^{i q_h^{\nu^\prime} a}} \end{array} \right),  \nonumber
\end{eqnarray}
where $t_N$ and $t_A$ are the normal and anomalous transmission coefficients, respectively,   while $q_e^\nu$ and $q_h^\nu$ are wave vectors in the bulk superconductor
\begin{equation}
q_{e(h)}^\nu =  \cos^{-1}\left(-\frac{\mu_{sc}+\epsilon_\nu\pm\sqrt{\omega^2-\Delta_0^2}}{2t_{sc}}\right),
\end{equation}
with $\epsilon_\nu = 2t_{sc}\cos[\nu \pi/(N_y+1)]$ and $|\omega|>\Delta_0$. 
To solve the BdG equation (\ref{SchEq}) with the boundary conditions (\ref{rcoeff}) and (\ref{tcoeff}) it is convenient to express the $16N_y$ wave function components $\Psi_{(j_x,j_y)}^{(\nu,\sigma)}$ corresponding to $j_x = 0, 1, N_{tot}, N_{tot}+1$ in terms of the $4N_y$ reflection and $4N_y$ transmission coefficients corresponding to the incoming channel $(\nu,\sigma)$. Thus, Eq. (\ref{SchEq}) reduces to a system of $4N_{tot}N_y$ linear equations with $4N_{tot}N_y$ unknown coefficients that include the $4(N_{tot}-2)N_y$ wave function components $\Psi_{(j_x,j_y)}^{(\nu,\sigma)}$, with $2\leq j_x \leq N_{tot}-1$, and the $8N_y$ reflection and transmission coefficients. Writing theses unknown coefficients as a vector $\widetilde{\Psi}$ with $4N_{tot}N_y$ components, we have 
\begin{equation}
\widetilde{\Psi}^{(\nu\sigma)}=({\cal H}+Q(\omega)-\omega)^{-1}J^{(\nu\sigma)}(\omega),
\label{GQ}
\end{equation} 

Explicitly, we define the components of the new vector  $\widetilde{\Psi}^{(\nu,\sigma)}$ to be the same as the wave function components, $\widetilde{\Psi}^{(\nu,\sigma)}_{(j_x,j_y)}\equiv\Psi^{(\nu,\sigma)}_{(j_x,j_y)}$, for $2\leq j_x \leq N_{tot}-1$, while
\begin{eqnarray}
\widetilde{\Psi}^{(\nu,\sigma)}_{1,j_y}&&\equiv
\sum_{\nu'}\phi_{\nu'}(j_y)
\begin{pmatrix} 
[r_N]_{\nu\sigma,\nu'\uparrow}e^{-ik^{\nu'}_e a}
\\
[r_N]_{\nu\sigma,\nu'\downarrow}e^{-ik^{\nu'}_e a}
\\
[r_A]_{\nu\sigma,\nu'\uparrow}e^{ik^{\nu'}_h a}
\\
[r_A]_{\nu\sigma,\nu'\downarrow}e^{ik^{\nu'}_h a}
\end{pmatrix},
\\
\widetilde{\Psi}^{(\nu,\sigma)}_{N_{tot},j_y}&&\equiv
\sum_{\nu'}\phi_{\nu'}(j_y) \nonumber
\\
\times&&\begin{pmatrix} \nonumber
u[t_N]_{\nu\sigma,\nu'\uparrow}e^{-iq^{\nu'}_e a}+v[t_A]_{\nu\sigma,\nu'\downarrow}e^{iq^{\nu'}_h a}
\\
u[t_N]_{\nu\sigma,\nu'\downarrow}e^{-iq^{\nu'}_e a}-v[t_A]_{\nu\sigma,\nu'\uparrow}e^{iq^{\nu'}_h a}
\\
u[t_A]_{\nu\sigma,\nu'\uparrow}e^{iq^{\nu'}_h a}-v[t_N]_{\nu\sigma,\nu'\downarrow}e^{-iq^{\nu'}_e a}
\\
u[t_A]_{\nu\sigma,\nu'\downarrow}e^{iq^{\nu'}_h a}+v[t_N]_{\nu\sigma,\nu'\uparrow}e^{-iq^{\nu'}_e a}
\end{pmatrix},
\end{eqnarray}
where $u$ and $v$ are BCS coherence factors:
\begin{equation}
u^2(\omega)=1-v^2(\omega)=\frac{1}{2}\left(1+\frac{\sqrt{\omega^2-\Delta^2}}{\omega}\right).
\end{equation} 
The matrix $Q$ introduces frequency-dependent correction at the boundaries,  
\begin{eqnarray}
\label{Q}
&&Q^{(\nu,\nu')}_{(1,j_y;~\!1,j_y')}(\omega)=t_{m}\delta_{\nu,\nu'}\delta_{j_y,j_y'} \nonumber 
\\
&&\times
\begin{pmatrix}
  e^{ik^{\nu}_e a} && 0 && 0 && 0
\\0 && e^{ik^{\nu}_e a} && 0 && 0
\\0 && 0 && -e^{-ik^{\nu}_h a} && 0
\\0 && 0 && 0 && 0 && -e^{-ik^{\nu}_h a}
\end{pmatrix},  
\\
&&Q^{(\nu,\nu')}_{(N_{tot};~\! j_y,N_{tot},j_y')}(\omega)=
\begin{pmatrix}
  Q^{(\nu,\nu')}_{ee}(\omega) && Q^{(\nu,\nu')}_{eh}(\omega)
\\-Q^{(\nu,\nu')}_{eh}(\omega) && Q^{(\nu,\nu')}_{hh}(\omega)
\end{pmatrix},  \nonumber
\end{eqnarray}
with
\begin{eqnarray}
&& Q^{(\nu,\nu')}_{ee}(\omega)= \frac{t_{sc}}{u^2-v^2}\delta_{\nu,\nu'}\delta_{j_y,j_y'} \nonumber
\\
&&\times
\begin{pmatrix}
  u^2 e^{iq^{\nu}_e a}-v^2e^{-iq^{\nu}_{h}a} && 0
\\0 && u^2 e^{iq^{\nu}_e a}-v^2e^{-iq^{\nu}_{h}a} 
\end{pmatrix},  \nonumber
\\
&& Q^{(\nu,\nu')}_{hh}(\omega)= \frac{t_{sc}}{u^2-v^2}\delta_{\nu,\nu'}\delta_{j_y,j_y'}\nonumber
\\
&&\times
\begin{pmatrix}
  u^2 e^{-iq^{\nu}_h a}-v^2e^{iq^{\nu}_{e}a} && 0
\\0 && u^2 e^{-iq^{\nu}_h a}-v^2e^{iq^{\nu}_{e}a} 
\end{pmatrix}, 
\\
&& Q^{(\nu,\nu')}_{eh}(\omega)= \frac{t_{sc}}{u^2-v^2}\delta_{\nu,\nu'}\delta_{j_y,j_y'} \nonumber
\\
&&\times
\begin{pmatrix}
  0 && uv(e^{-iq^{\nu}_h a}- e^{iq^{\nu}_{e}a})
\\uv(e^{-iq^{\nu}_h a}- e^{iq^{\nu}_{e}a}) && 0
\end{pmatrix}.  \nonumber
\end{eqnarray}
All other elements of $Q$ are zero,  $Q^{(\nu,\nu')}_{(j_x,j_y;~\! j_x',j_y')}(\omega)=0$ if $j_x\neq j_x^\prime$ or $j_x =j_x^\prime \neq 1, N_{tot}$. Finally, the vector $J$ contains the ``free terms'' from the boundary conditions (i.e., those that do not depend on the reflection and transmission coefficients), which are associated with the incoming plane wave. Explicitly, we have 
\begin{eqnarray}
J^{(\nu\sigma)}_{(1,j_y)}(\omega)=&&-t_{m}+(\omega-\mu_{m})e^{i k^{\nu}_e a}, \nonumber
\\
J^{(\nu\sigma)}_{(2,j_y)}(\omega)=&&-t_m e^{i k^{\nu}_e a}, 
\end{eqnarray}
and $J^{(\nu\sigma)}_{(j_x,j_y)}(\omega)=0$ for $j_x\geq 3$.
Note that all the information about the boundary conditions is contained in the matrix $Q(\omega)$ and the  vector $J^{(\nu\sigma)}(\omega)$. The coefficients $[r_{N(A)}]_{\nu\sigma,\nu^\prime\sigma^\prime}$ and $[t_{N(A)}]_{\nu\sigma,\nu^\prime\sigma^\prime}$ are obtained by solving Eq. (\ref{GQ}) $2N_y$ times, once for each incoming channel. 

An interesting observation is that the matrix in Eq. (\ref{GQ}) has the structure of a Green's function, $G_Q(\omega)  = ({\cal H}+Q(\omega)-\omega)^{-1}$. Generalizing the approach used in Sec. \ref{S_IIA} to calculate the effective Green's function for the semiconductor wire, we integrate out the degrees of freedom of the bulk superconductor and write an equation for the reflection coefficients involving  only states in the semiconductor and the metallic lead: 
\begin{equation}
\widetilde{\Psi}^{\prime(\nu\sigma)}=({\cal H}_{m}+{\cal H}_{sm}+{\cal T}_m+\Sigma_{Q}+Q_{m}-\omega)^{-1}J^{\prime(\nu\sigma)}(\omega), \label{GQ1}
\end{equation}  
where ${\cal H}_{m}$, ${\cal H}_{sm}$, and ${\cal T}_m$ are the matrices corresponding to the first quantized Hamiltonians for the metallic lead, the semiconductor wire, and the lead-wire coupling, respectively, $Q_{m}$ is the matrix corresponding to the boundary condition at the left end of the normal lead and $\Sigma_{Q}$ is a ``self-energy'' contribution that incorporates the effect of the parent superconductor, including the the outgoing boundary conditions. Note that below the bulk gap there is no normal current (i.e., current carried by quasiparticles) in the superconductor; hence the outgoing boundary conditions become trivial and we have $\Sigma_{Q}(\omega) = \Sigma_{sc}(\omega)$. In general, since the boundary conditions are diagonal in real space while $H_{sc}$ is diagonal in mode space, the self energy $\Sigma_Q$ has to be calculated numerically. Nonetheless, using Eq. (\ref{GQ1}) instead of Eq. (\ref{GQ}) provides significant advantages, as one can calculate the self-energy $\Sigma_Q$ once, then solve the reduced equation repeatedly for different values of relevant parameters, such as potential barrier height and Zeeman field, instead of solving the full equation  (\ref{GQ})  for each set of  parameters. 

The differential conductance for tunneling into the hybrid structure can be written in terms of reflection coefficients as\cite{Blonder1982} 
\begin{equation}
\frac{dI}{dV}=\frac{e^2}{h}\!\sum_{\nu,\nu^\prime}\sum_{\sigma,\sigma^\prime}\left(1\!-\!|[r_N(V)]_{\nu\sigma,\nu^\prime\sigma^\prime}|^2\!+\!|[r_A(V)]_{\nu\sigma,\nu^\prime\sigma^\prime}|^2\right), \label{dIdV}
\end{equation}
where $V$ is the bias voltage. To include the effect of finite temperature, the conductance is broadened by convolving it with the Fermi function. Explicitly, we have
\begin{equation}
G(V,T)=\int d\epsilon\frac{G_0(\epsilon)}{4T\cosh^2\left(\frac{V-\epsilon}{2k_BT}\right)}
\end{equation}
where $G_0(\epsilon)=dI/dV$ is the zero temperature conductance given by Eq. (\ref{dIdV}) at voltage bias $\epsilon$.

\subsection{Keldysh Formalism} \label{S_IIC}

Consider a normal metal-semiconductor junction. The current through the junction can be expressed in terms of the number operator for the  component to the right of the junction as
\begin{equation}
I=e \bra{\Psi_0}\dot{N}_R\ket{\Psi_0}=\frac{-i e}{\hbar}[N_R,H],
\end{equation}
where $\ket{\Psi_0}$ is the equilibrium quantum state of the composite system, $H$ is the total Hamiltonian, and $N_R=\sum_r c^{\dagger}(r)c(r)$ is the number operator for the right component. The label $r$ indicates both position (on the right side of the junction) and spin. The number operator $N_R$ can only change if some electrons cross the junction. Hence, one can rewrite the current as\cite{Berthod2011}
\begin{equation}
\label{current_a}
I=\frac{-2 e}{\hbar}\sum_{l,r}[T_m]_{lr} {\rm Im}[\bra{\Psi_0}{\rm a}^{\dagger}(l)c(r)\ket{\Psi_0}]
\end{equation}
where $T_m$ is the coupling between the metal and the semiconductor given by Eq. (\ref{msm}).  The differential conductance can be extracted by taking the derivative of the current  with respect to the bias voltage in the lead, $\frac{dI}{dV}$. Notice that the propagator $iG^<(l,r;0^+) = \bra{\Psi_0}\rm{a}^{\dagger}(l)c(r)\ket{\Psi_0}$ is the so-called lesser Green's function at time $t=0^+$.\cite{Rammer1986}  Using the relation between $G^<(l,r;t)$ and the Keldysh Green's function, Eq. (\ref{current_a}) becomes
\begin{equation}
\begin{split}
I&=-\frac{2e}{\hbar}\sum_{l,r}[T_m]_{lr} {\rm Re}[G^<(l,r;0^+)]
\\
&=\frac{e}{\hbar}\int d\omega~\! {\rm Re} {\rm Tr}\left[T_m G^K(\omega)\right],
\end{split}
\end{equation}
where we have used the property  $\int d\omega G^K(\omega)=G^K(l,r;0^+)$.  
Expanding $G^K$ in terms of Green's functions for the left and right sub-systems, we have 
\begin{eqnarray}
G &\equiv&
\begin{pmatrix}
G^+ & G^K \\ 0 & G^-
\end{pmatrix} \nonumber \\
&=& G_R T_m G_L+G_R T_m G_L T_m G_R T_m G_L+\dots
\end{eqnarray}
The Keldysh Green's function becomes
\begin{eqnarray}
G^K &=& [1-G^+_R T_m G^+_L T_m]^{-1} \\
&\times & (G^+_R T_m G^K_L+G^K_R T_m G^-_L)[1-G^-_R T_m G^-_L T_m ]^{-1}, \nonumber
\end{eqnarray}
where $G^+$ and $G^-$ are the retarded and advanced Green's functions, respectively.
Next, we take the derivative with respect to the bias voltage.  We assume the voltage drops at the junction, so we only take the derivative in the Green's function of the lead, $G_L$ (i.e., we assume that $G_R$ is independent of $V$).  The main contribution comes from the derivative of the Fermi function with respect to the voltage, hence from $G^K_L(\omega)=[1-2f_{L}(\omega)][G^+_L(\omega)-G^-_L(\omega)]$, where $f_L(\omega) = f(\omega-eV)$. Explicitly, the differential conductance is given by  
\begin{eqnarray}
\frac{dI}{dV} &=& -\frac{2e^2}{\hbar}\int d\omega\frac{df(\omega-eV)}{d(eV)} \nonumber \\
&\times & {\rm Re}\left\{{\rm Tr}\left[ T_m \left(1-G^+_R T_m G^+_L T_m\right)^{-1} G^+_R T_m \right.\right. \\
&\times & \left.\left.\left(G^-_L-G^+_L\right)\left(1-G^-_R T_m G^-_L T_m\right)^{-1}\right]\right\} \nonumber
\end{eqnarray}
Finite temperature can be included in the same way as in the BTK method.

\section{Results} \label{S_III}

In this section we discuss the results of our numerical analysis focusing on features in the calculated differential conductance that are only present if the parent superconductor is explicitly treated as an active component of the system and on the dependence of the Majorana zero bias conductance peak on the relevant system parameters.

The numerical results described below are obtained by setting the chemical potentials for the metal and the superconductor ($\mu_{m}$ and  $\mu_{sc}$, respectively) to the middle of their respective bands. The other model parameters are chosen as follows: 
 The lattice constant is $10~$nm along the wire (i.e., in the $x$ direction) and $33~$nm across the wire (i.e., in the $y$ direction); if the wire is $1~\mu$m long, there are 100 sites in each SM chain.  The hopping parameters are $t_{sm}^x=9.5~$meV, $t_{sm}^y=0.95~$meV, $t_{m}^x==3.8~$meV, $t_{m}^y=0.38~$meV, $t_{sc}^x=1.5~$meV, $t_{sc}^y=0.15~$meV, and $\tilde{t}_{m-sm}=3.0~$meV.  The spin orbit coupling parameters are $\alpha_x=2~$meV and $\alpha_y=0.6~$meV.  
 The bias potential is applied to the metallic lead, which is $40$ sites long (the shortest length that is consistent with plane wave boundary conditions), while the superconductor is grounded. 
 
 We model the parent superconductor as a long, thick slab having $1000$ sites in the $x$ direction (i.e.,  $10~\mu$m) and $300$ sites in the $z$ direction. To handle the problem concerning the large number of degrees of freedom associated with the parent superconductor, we calculate the (surface) Green's function of a large slab {\em once} and store the information. Subsequently, we ``attach'' this large superconductor to semiconductor wires (of various lengths) and calculate the corresponding interface self-energy, which completely contains the effect of the superconductor.
 More specifically, we first calculate numerically the superconductor Green's function, which determines both the self energy $\Sigma_{sc}$ given by (\ref{Sig_sc}) and the boundary-condition-dependent ``self-energy'' $\Sigma_Q$ from Eq. (\ref{GQ1}), in the transverse mode space,  where the problem is diagonal and can be solved separately for each of the modes. Then, we calculate the surface Green's function for frequencies up to $3~$meV taking $100$ equally spaced values in the range $2$ to $3~$meV. If necessary, we extrapolate for frequencies in-between these these discrete values. Note that below the bulk gap (i.e., below $2~$meV), we can use the analytical solution given by Eq. (\ref{fullscse}).  These calculated values of the surface Green's function are stored for later use in the calculation of the self-energies corresponding to wires of different lengths.


\subsection{Method comparison and differential conductance vs. LDOS}  \label{S_IIIA}

\begin{figure}[t]
\begin{center}
\includegraphics[width=0.48\textwidth]{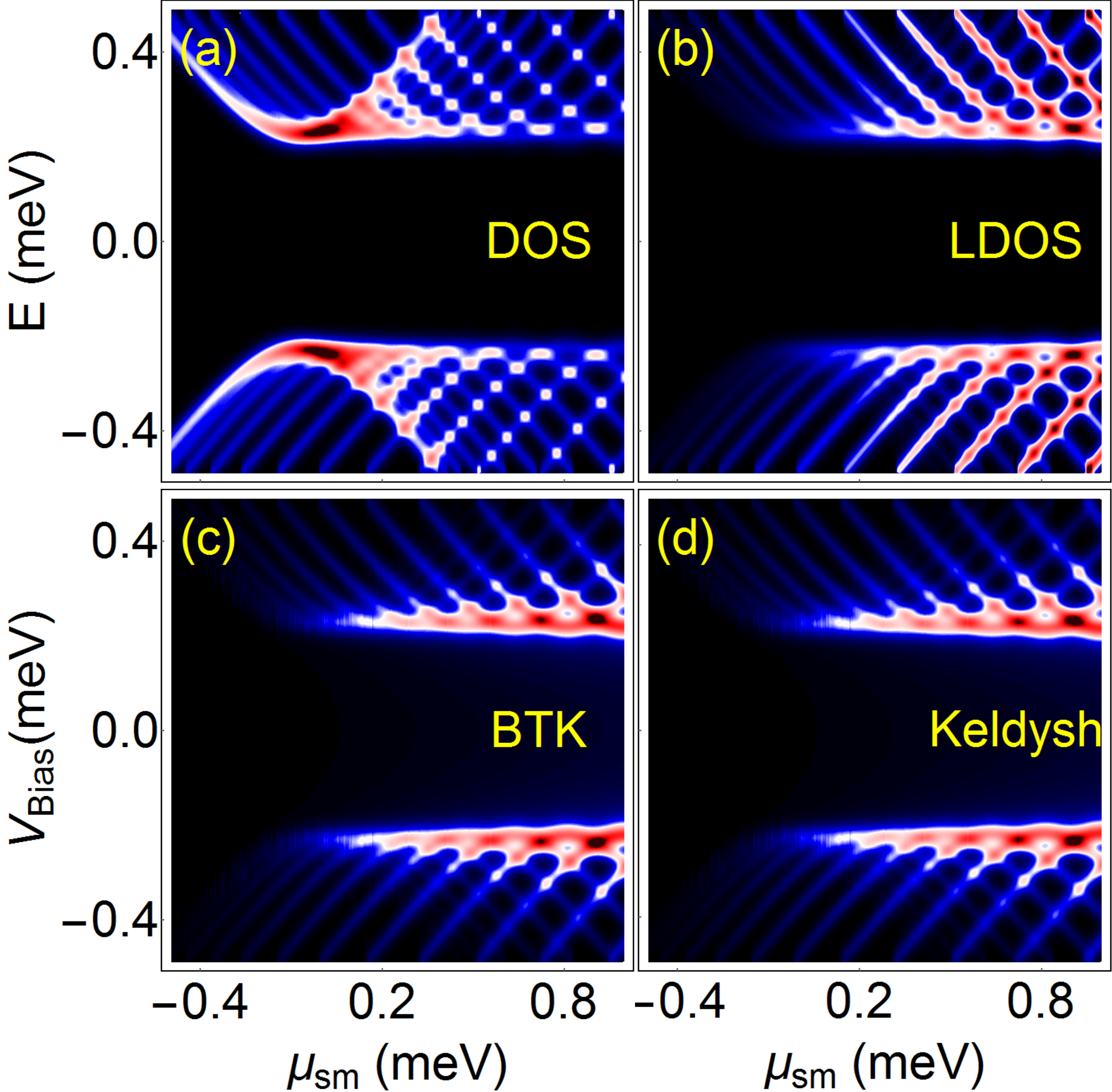}
\end{center}
\vspace{-2mm}

\caption{(Color online) Differential conductance ($dI/dV$), density of states (DOS), and local density of states (LDOS) at the end of the proximitized wire adjacent to the tunnel barrier as functions of the chemical potential and bias voltage/energy. The differential conductance is calculated using both the BTK [panel (c)] and Keldysh [panel (d)] methods. Note that $dI/dV$ reflects (qualitatively) the local density of states near the barrier [panel (b)], rather than the total DOS [panel (a)]. The length of the wire is $L=1~\mu$m, the parent superconductor gap is $\Delta_0= 2~$meV, and the induced gap is $\Delta_{ind}=0.25~$meV.  The Zeeman field is turned off.}
\label{comp}
\vspace{-3mm}
\end{figure}

The BTK\cite{Blonder1982} and Keldysh\cite{Rammer1986} formalisms are two possible methods for calculating the differential conductance.  Since the two approaches are formally rather different, the first question that we want to address is whether or not the predictions based on these methods are the same. In addition, when interpreting charge transport measurements on semiconductor -superconductor (SM-SC) Majorana hybrid structures it is essential to understand the connection between the measured differential conductance and the low-energy spectral properties of the hybrid system. To address these questions, we calculate the density of states (DOS) and the local density of states (LDOS) at the end of the wire for a nanowire of length $L=1~\mu$m proximity-coupled to a $\Delta_0=2~$meV superconductor so that the induced gap is $\Delta_{ind}=0.25~$meV.  We compare these quantities with the differential conductance for tunneling into the end of the same wire, which is calculated using both methods described in the previous section. The results are shown in Fig. \ref{comp}.

Panels (a) and (b) show the DOS and LDOS, respectively, as functions of the chemical potential and energy. The bottom panels in Fig. \ref{comp}, i.e., (c) and (d), show the differential conductance calculated using the BTK and Keldysh methods.
The ``stripy'' features are due to the quantization of states in a finite wire and  reflect the variation of the eigenvalues of the BdG Hamiltonian with the chemical potential. 
In our model the energy difference between successive states becomes larger as the chemical potential increases.
Note that  $\mu_{sm}=0$ corresponds to the bottom of the band, which is clearly revealed by the large DOS in panel (a). By contrast, there is no clear signature associated with the bottom of the band in the LDOS [panel (b)] or in the differential conductance [panels (c) and (d)]. This is due to the fact that the LDOS is calculated at the end of the proximitized wire adjacent to the potential barrier, where states from the bottom of the band have very low amplitude.  These states are weakly coupled to the lead and, consequently, do not contribute significantly to the differential conductance, as shown in panels (c) and (d). Another way of understanding this behavior is in terms of velocity.
Penetration of the barrier is determined by the group velocity.  An electron occupying a state with higher velocity has a greater chance to get through the barrier and couple to the lead.  Since the velocity vanishes at the bottom of the band, the corresponding states will have a small contribution to the conductance.  As the chemical potential increases, the wavelength of the states is shortened and their velocity becomes larger, which results  in higher values of the conductance.

Based on this analysis we conclude that (i) the BTK and Keldysh formalisms predict similar values for the differential conductance and (ii) the dependence of the differential conductance on the chemical potential and bias voltage is qualitatively similar to that of the LDOS at the end of the wire adjacent to the tunnel barrier. Of course, the second property holds in the weak tunneling regime,\cite{Berthod2011} but breaks down when the transparency of the barrier is large. In Fig. \ref{comp} the only notable difference between the differential conductance and the LDOS is the decrease of $dI/dV$ with increasing bias voltage $|V_{Bias}|$, which is not reflected in the dependence of the LDOS on energy. To understand this behavior, we note that for voltage values below the gap of the parent superconductor charge can only travel through the superconductor as a supercurrent.  Tunneling electrons into the wire involves anomalous (Andreev) reflection processes\cite{Blonder1982,He2014} (an incoming electron returns as a hole) and the probability for such a process depends on the properties of the quasiparticle states of the wire-superconductor hybrid system.  Near the induced gap ($\Delta_{ind}\approx 0.25$ meV in Fig. \ref{comp}), where every state is an almost equal blend of particles and holes, the probability of anomalous reflection is large, but it decreases at higher (in absolute value) energies, where the quasiparticles have a dominant particle or hole character.    

We have verified our conclusions for a broad range of parameters, including bias voltage values larger than the parent superconductor gap, and we found them to be rather generic properties. Below we calculate the conductance using the most convenient method for a specific setup or physical aspect that we want to investigate. Since we are mostly interested in the weak tunneling regime, we will interpret the results as (approximately) representing the local density of states near the end of the wire-superconductor system adjacent to the tunnel barrier.

\subsection{Dependence on the semiconductor-superconductor coupling strength and tunnel barrier height}  \label{S_IIIB}

Proximity coupling a semiconductor wire to a conventional superconductor results in the opening of an induced gap in the quasiparticle spectrum at the chemical potential. At zero momentum, $k=0$, and in the absence of a magnetic field the induced gap $\Delta_{ind}$ is determined by the gap $\Delta_0$ of the parent superconductor and by the effective semiconductor-superconductor (SM-SC) coupling constant $\gamma$.  The simplest way to account for the emergence of an induced gap is to introduce a pairing potential in the effective BdG Hamiltonian  for the wire. However, this approach fails to describe accurately the low-energy physics of the hybrid structure (i.e., the induced "bulk" quasiparticle gaps and the in-gap bound states), with the exception of the weak coupling limit $\gamma/\Delta_0 \rightarrow 0$. However, experimentally-relevant devices are not in this limit and the explicit treatment of the parent superconductor, e.g., using  Eq. (\ref{fullscse}), is necessary.\cite{Stanescu2017a} What features of the differential conductance are directly related to the parent superconductor being an active component of the hybrid structure?

\begin{figure}[t]
\begin{center}
\includegraphics[width=0.48\textwidth]{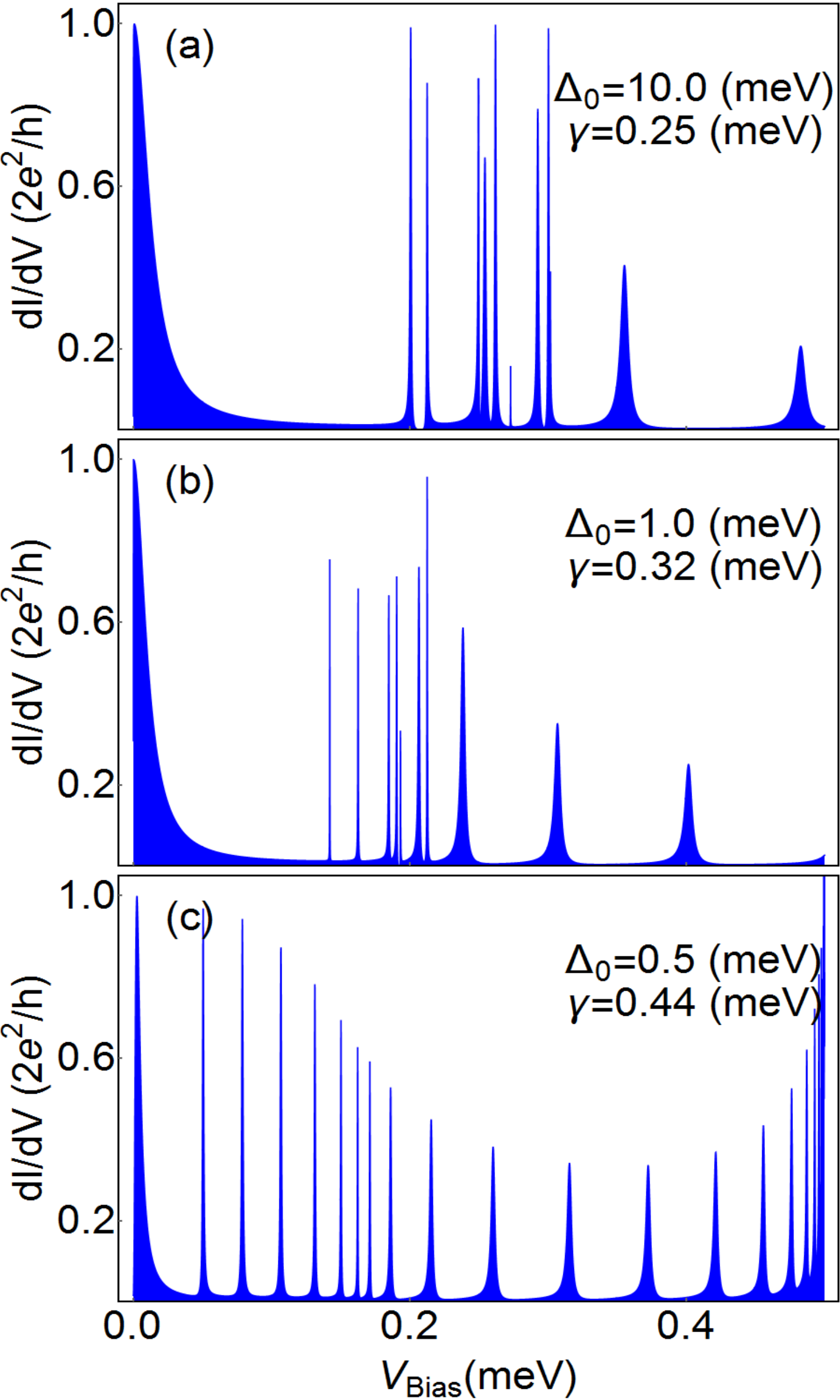}
\end{center}
\vspace{0.mm}

\caption{(Color online) Differential conductance as a function of bias potential for a wire of length  $L=1$ $\mu$m in the presence of  a finite magnetic field $\Gamma=0.5$ $meV$ for different ratios $\gamma/\Delta_0$ that correspond to a constant induced gap 
$\Delta_{ind}\approx 0.25$ meV. (a) Weak coupling regime with $\gamma=0.25$ meV and $\Delta_0=10$ meV. The conductance curve is nearly identical with that calculated based on the effective pairing approximation, Eq. \ref{Din}. (b) System with $\Delta_0= 1$ meV and  $\gamma=0.32$ meV. (c) $\Delta_0=0.5$ meV and  $\gamma=0.44$ meV.  In the lower panels the location of the peaks is shifted and the weight of each peak is reduced, revealing the effects of the proximity-induced low-energy renormalization.}
\label{D0}
\vspace{-3mm}
\end{figure}

First, we consider the dependence of the low-energy features of a finite wire in the topological regime on the effective SM-SC coupling strength. The differential conductance as function the bias voltage for three different ratios $\gamma/\Delta_0$ is shown in  Fig. \ref{D0}. The calculations are done for a wire of length $L=1~\mu$m in the presence of a Zeeman field $\Gamma=0.5~$meV. Note that the induced gap (i.e., the quasiparticle gap at $\Gamma=0$) is the same in all three cases, $\Delta_{ind}=0.25~$meV. However, the critical Zeeman splitting corresponding to the topological quantum phase transition (TQPT) is different, as it is determined by the effective coupling strength $\gamma$, rather than the induced gap.\cite{Stanescu2017a,Sticlet2016} For the Zeeman field value used in the calculations, $\Gamma=0.5~$meV, all three systems are in the topological regime, as signaled by the presence of a Majorana-induced zero-bias conductance peak.
The differential conductance in Fig. \ref{D0}(a), which corresponds to $\gamma\ll\Delta_0$, is the same as that predicted by the effective pairing approximation defined by Eq. \ref{Din}.  However, increasing the effective coupling [panels (b) and (c)] results in  several effects that cannot be captured by this approximation. First, the weight of the Majorana-induced zero-bias peak (ZBP) decreases with increasing coupling. The main reason for this behavior is the fact that the spectral weight of the Majorana mode within the wire decreases with increasing coupling. In other words, as $\gamma$ increases, the amplitude of the Majorana wave function within the wire (i.e., in the vicinity of the tunnel barrier) is reduced, while its amplitude within the parent superconductor (i.e., farther away from the tunnel barrier) is enhanced.\cite{Stanescu2017a} This results in a lower ``visibility'' of the Majorana mode, as revealed by the ZBP. The second effect is the renormalization of the quasiparticle energies and the reduction of the topological gap.\cite{Stanescu2017a} In essence, these effects are due to the fact that all low-energy states reside both in the wire {\em and} in the parent superconductor, with more and more spectral weight being transfered to the superconductor as the coupling strength increases.  

The bias voltage range shown in Fig. \ref{D0} corresponds to energies below the gap of the parent superconductor. In this regime there is no normal current flowing through the superconductor. In the BTK approach, this means that it is sufficient to impose boundary conditions at the end of the normal lead.  However, if we want to explore an energy window larger than the bulk gap, we have to supplement this with boundary conditions at the end of the parent superconductor.  These boundary conditions account for the normal current carried by the quasiparticles in the parent superconductor. Note that these boundary conditions are diagonal in real space, while the superconductor Hamiltonian is diagonal in the mode space.  Consequently, finding an analytical expression is rather difficult;  instead, we calculate  the self-energy $\Sigma_{sc}(\omega)$  (which incorporates the boundary conditions) numerically.     

\begin{figure}[t]
\begin{center}
\includegraphics[width=0.48\textwidth]{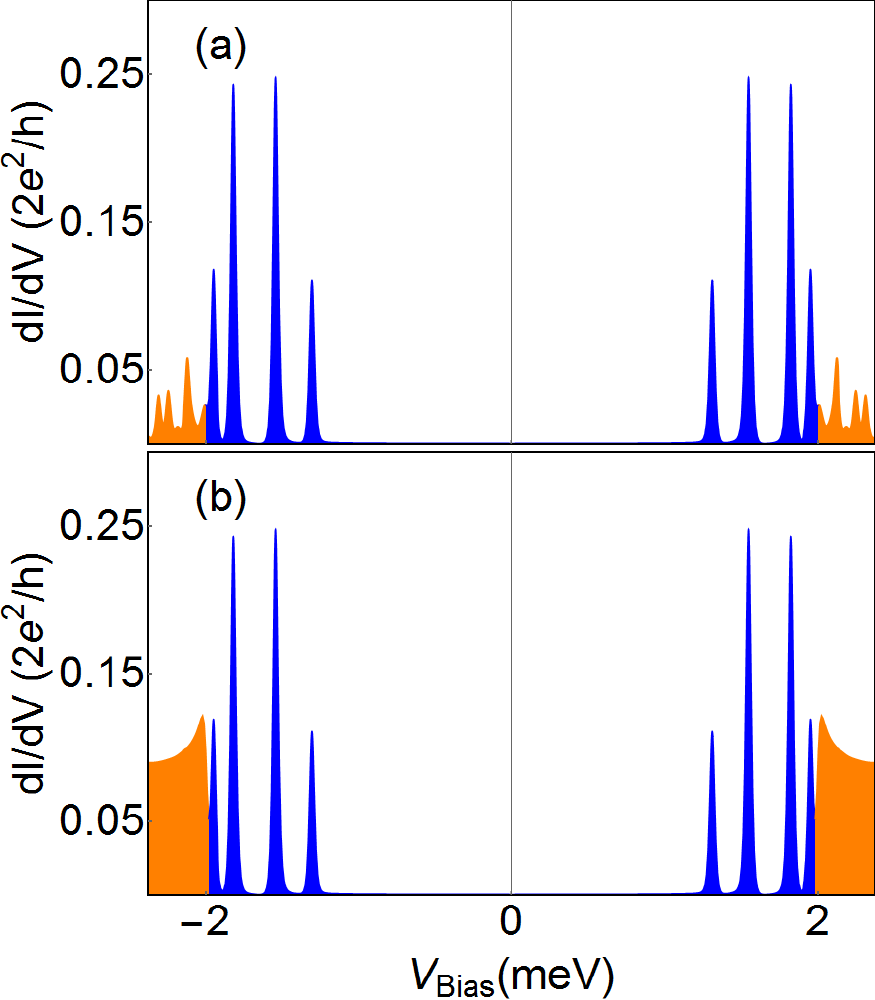}
\end{center}
\vspace{-3mm}

\caption{(Color online) (a) Differential conductance calculated using the BTK formalism and including only anomalous reflection processes. (b) Differential conductance that includes both Andreev reflection processes and contributions from the normal transmission through the parent superconductor.  Blue (darker gray) represents features below the bulk superconductor gap, while orange (light gray) marks contributions above the gap. The model parameters are:  $L=0.3~\mu$m, $\Gamma=0$, $\Delta_0=2~$meV,  $\gamma=1.5~$meV, and $k_BT=0.002~$meV.}
\label{scnsc}
\vspace{-3mm}
\end{figure}

Including the normal transmission is required at bias voltages above the bulk gap, as demonstrated by the data shown in Fig. \ref{scnsc}.  Here we calculate the differential conductance for a short  nanowire of length $L=0.3~\mu$m coupled to a superconductor with $\Delta_0=2~$meV. The effective semiconductor-superconductor coupling is $\gamma=1.5~$meV.  In the top panel, the differential conductance is calculated without including the contribution from the normal transmission through the parent SC.  All the features in this panel represent anomalous current contributions, which become rather small at energies larger than the bulk gap. Note that there is no feature associated with the bulk gap edge. By contrast,  in the bottom panel, which shows the full calculation that includes the normal current contribution, one can clearly observe an additional feature associated with the parent superconductor gap (orange/light gray area). The emergence of an additional conductance peak at the energy corresponding to the gap $\Delta_0$ of the parent superconductor, which was recently discussed in Ref. \onlinecite{Reeg2017}, confirms the the predictions of Ref. \onlinecite{Cole2015}, which are based on a local density of states analysis. The presence of features associated with the parent superconductor may generate certain difficulties when interpreting $dI/dV$ measurements for hybrid systems in the intermediate and strong coupling regimes. In these cases, the induced gap $\Delta_{ind}$ and the bulk gap of the parent superconductor $\Delta_0$ have comparable values and, therefore, it is important to be able to distinguish the corresponding features.  To this end, we analyze their dependence on the effective SM-SC coupling strength and on the tunnel barrier height.

\begin{figure}[t]
\begin{center}
\includegraphics[width=0.47\textwidth]{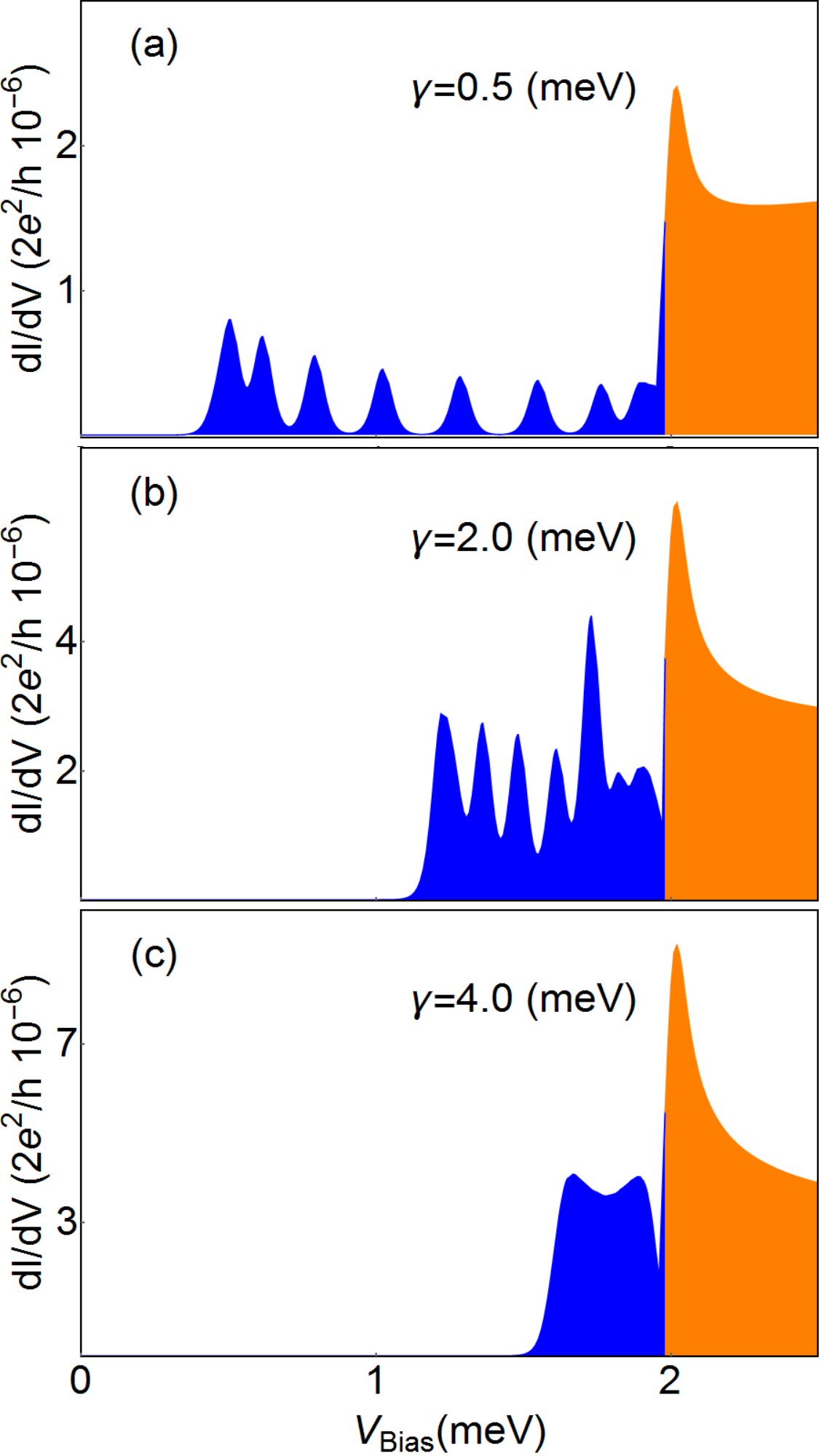}
\end{center}
\vspace{-3.5mm}

\caption{(Color online) Differential conductance in the weak tunneling limit as a function of the bias voltage for different values of the effective SM-SC coupling $\gamma$. The model parameters are $L=1~\mu$m, $\Delta_0 = 2~$meV, $\Gamma=0$, and $k_BT=0.005~$meV. The induced gap increases with $\gamma$, approaching $\Delta_0$ in the strong coupling limit $\gamma/\Delta_0\rightarrow\infty$, while the spacing between  energy eigenstates decreases. The conductance peak associated with the bulk gap edge (orange/light gray) becomes more pronounced in the strong coupling regime.}
\label{gamma}
\vspace{-3mm}
\end{figure}

As discussed in the relation to Fig. \ref{D0}, increasing the coupling between the semiconductor wire and the superconductor results in a renormalization of the low-energy spectrum and a reduction of the spectral weight within the wire. In addition, varying $\gamma$ has a strong impact on the induced gap $\Delta_{ind}$. In a multi-band system the specific dependence of the induced gap on the SM-SC coupling strength is rather complicated (except in the weak coupling regime) and involves additional parameters, such as the inter-band spacing and chemical potential.\cite{Stanescu2017a} On the other hand, the feature in the differential conductance associated with the bulk superconductor has a relatively weak dependence on the SM-SC coupling, as shown in Fig. \ref{gamma}.  In this figure, we calculate  again at the differential conductance for a nanowire coupled to a  superconductor with $\Delta_0=2~$meV, but the length of the nanowire is now $L=1~\mu$m; the effective SM-SC coupling takes three different values, $\gamma = 0.5, ~2.0,$ and $4.0~$meV.  Note, that the peak associated with the bulk gap edge becomes stronger with increasing $\gamma$. The features corresponding to low-energy semiconductor states (blue/darker gray) reveal the dependence of the induced gap on the SM-SC coupling and the proximity-induced energy renormalization discussed above. In particular, the energy spacing between successive low-energy states decreases with $\gamma$, so that a finite system characterized by discrete spectral features in the low-coupling regime [panel (a)] appears as having a continuous spectrum for a high-enough coupling strength [panel (c)]. 

\begin{figure}[t]
\begin{center}
\includegraphics[width=0.48\textwidth]{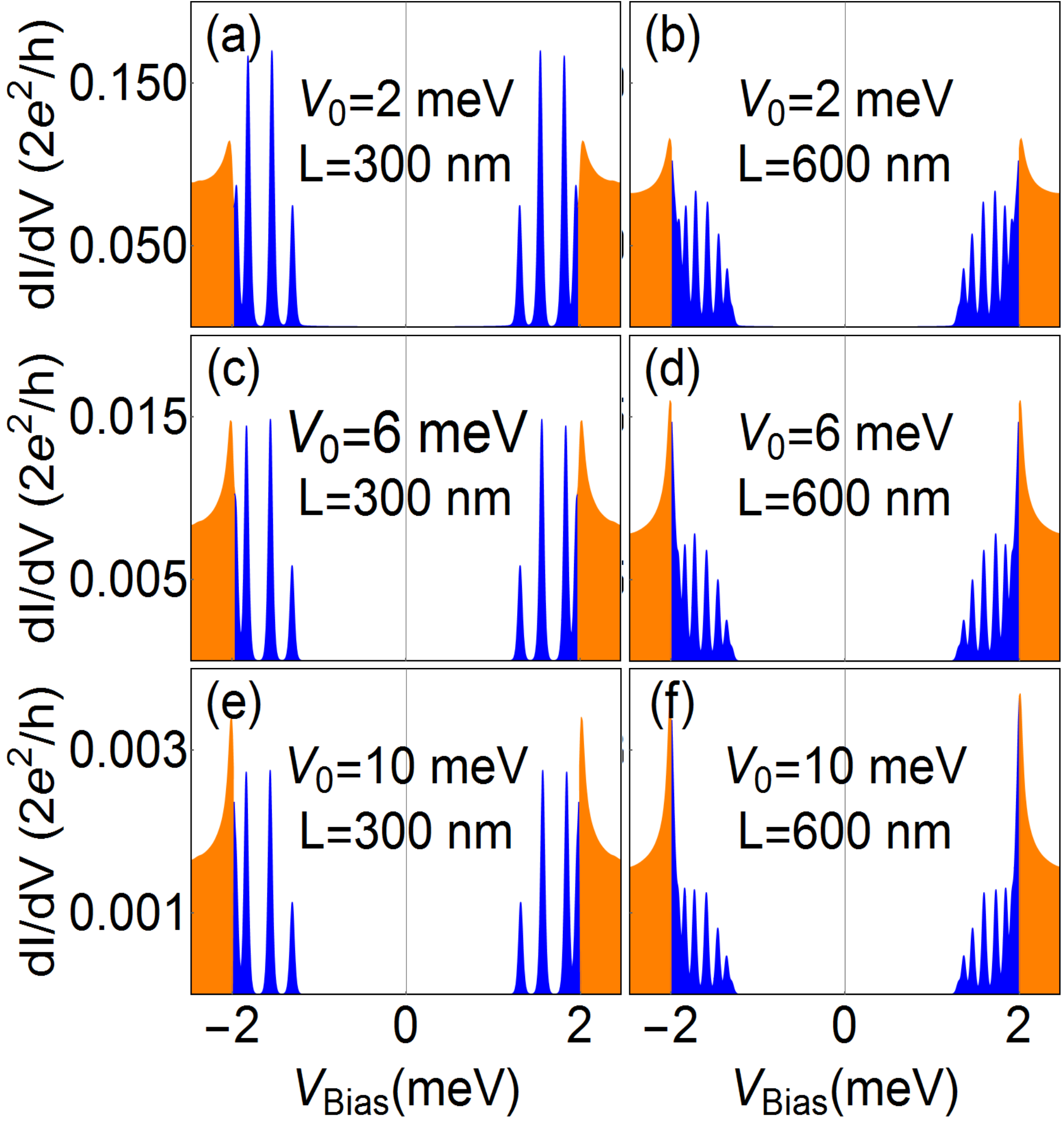}
\end{center}
\vspace{-3.5mm}

\caption{(Color online) Differential conductance as a function of the bias voltage for different values of the  potential barrier and wire length.  Increasing the length of the semiconductor wire increases the number of sub-gap states (blue/darker gray), but does not affect the feature associated with the parent superconductor (orange/light gray). Increasing the potential barrier suppresses the differential conductance, but the sub-gap features are significantly more affected than the bulk contribution.  The model parameters are:  $\Gamma=0$, $\Delta_0=2~$meV,  $\gamma=1.5~$meV, and $k_BT=0.002~$meV.}
\label{V0}
\vspace{-3mm}
\end{figure}

\begin{figure}[t]
\begin{center}
\includegraphics[width=0.48\textwidth]{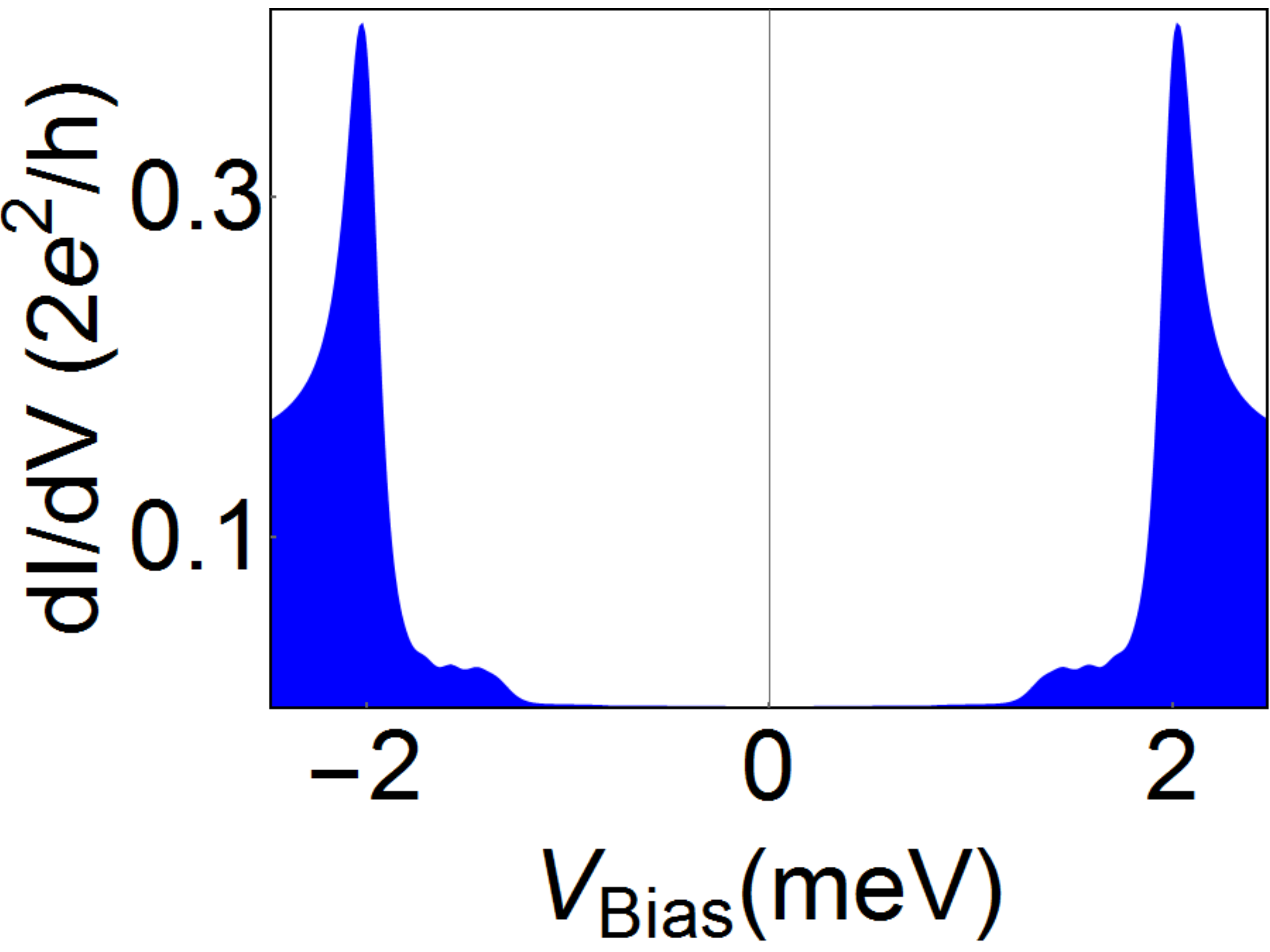}
\end{center}
\vspace{-2mm}

\caption{(Color online)  Differential conductance as function of the bias voltage for a nanowire with $L=0.6~\mu$m (also see Fig. \ref{V0}, right panels) and  a barrier potential that penetrates deeper under the superconductor.   Note the substantial difference between the conductance above the (bulk SC) gap and the sub-gap conductance.  The model parameters are:  $\Gamma=0$, $\Delta_0=2~$meV,  $\gamma=1.5~$meV, and $k_BT=0.004~$meV.}
\label{suppress}
\vspace{-3mm}
\end{figure}

The dependence of the differential conductance on the tunnel barrier height is illustrated by the data shown in Fig. \ref{V0}. We calculate $dI/dV$ as a function of the bias voltage for different barrier heights and two different wire lengths $L=0.3~\mu$m and $L=0.6~\mu$m.  As before, the parent SC gap is $\Delta_0=2~$meV, while the SM-SC coupling is $\gamma=1.5~$meV. First, we note that the sub-gap features associated with the presence of the proximitized semiconductor wire (blue/darker gray) are strongly dependent on the length of the structure. By contrast, the parent superconductor contribution (orange/light gray) is practically independent on $L$. In very short wires [panels (a), (c), and (e)] finite size quantization results in a discrete spectrum that generates a set of well-separated peaks in the differential conductance. Increasing the length of the wire  [panels (b), (d), and (f)] reduces the intervals between successive peaks and, eventually, the spectrum becomes continuous [see, for example, Fig. \ref{gamma} (c) and Fig. \ref{suppress}]. Note that the energy separation required for resolving discrete peaks depends on  temperature, barrier transparency, and strength of the disorder present in the system. 
 Above the bulk gap of the parent superconductor the semiconductor states hybridize with the superconductor states, which are overwhelmingly more numerous and, practically, determine the conductance response.  Consequently, above $\Delta_0$ the conductance is nearly independent of the wire  length. Note, however, that the semiconductor plays an important role here. One can view all the hybrid states above the gap as superconductor states that have acquired a small ``tail'' that extends into the wire, including the region adjacent to the tunnel barrier. While the spectral weight carried by the tails is negligible,  they are essential for ensuring the coupling between the normal lead and the ``superconductor states.'' In other words, the orange/light gray component of the differential conductance corresponds to hybrid states that have most of their spectral weight inside the parent superconductor, but have small tails extending into the barrier region.  

Varying the strength of the barrier potential affects the differential conductance both  below and above the bulk gap (see Fig. \ref{V0}).  However, the differential conductance below the bulk gap (blue/darker gray) decreases faster than the conductance above the gap (orange/light gray) with increasing barrier potential.  In fact, it is possible that for a large enough barrier potential the sub-gap conductance is ``completely'' suppressed (for a certain finite resolution), while the  parent superconductor contribution is still measurable.  
A rigorous proof of this statement would imply performing a (self-consistent) calculation of the barrier potential profile, i.e., finding a (self-consistent) solution of the Schr\"odinger-Poisson equation. Intuitively, one can understand the effect of the barrier potential as resulting from the huge difference between the carrier densities in the semiconductor and the superconductor. Increasing the potential of the tunneling gate results in the potential barrier penetrating into the proximitized segment of the nanowire (i.e., under the superconductor) and ``pushing away'' the low-energy states that reside inside this component of the hybrid structure. On the other hand, the effect on the superconductor is negligible due to the screening provided by the large carrier density.  To illustrate this effect, we calculate the differential conductance for a nanowire with $L=0.6~\mu$m, similar to that corresponding to the right panels in Fig. \ref{V0}, and  a barrier potential profile that penetrates more into the wire. The results are shown in  Fig. \ref{suppress}.  Note the large difference between the conductance above the (bulk SC) gap and the sub-gap conductance, which appears as a small tail of the coherence peak. Remarkably, this shape looks quite similar to recent experimental results on epitaxial  Al-InAs superconductor-semiconductor nanowires\cite{Chang2015} (more specifically, Fig. 5(c) for the half-shell device). To experimentally disentangle the induced and bulk contributions in the strong SM-SC coupling limit (where they have similar energy scales) we propose an analysis of the dependence of various features on the  barrier potential.  Starting, for example, with an extremely high barrier will only reveal bulk-type contributions.The induced features will progressively emerge as the potential barrier is lowered.

\subsection{Parent superconductor with sub-gap states}  \label{S_IIIC}

So far we have assumed an ``ideal''  parent superconductor characterized by a perfectly clean gap, even in the presence of finite magnetic fields. However, real superconductors may have sub-gap states  induced by disorder and finite external fields. Nonetheless, if a ``dirty'' superconductor has
a few localized sub-gap states, we expect the system to behave qualitatively similar to a clean system. Indeed, the additional sub-gap states may hybridize with low-energy states from the semiconductor wire, which will result in the hybrid structure having additional low-energy modes and smaller quasiparticle gaps. However, the structure of the low-energy spectrum will be qualitatively similar to that of a clean system. Moreover, if the sub-gap states are localized far from the ends of the wire the Majorana modes will not be affected. The situation is completely different if the parent superconductor has a small but finite density of sub-gap states. In this situation, the entire gap is populated with hybrid states having most of their spectral weight inside the parent  superconductor and some of it inside the semiconductor nanowire.\cite{Stanescu2017a} The question that we want to address is the following: what are the characteristic signature in the tunneling conductance of a hybrid structure with a parent superconductor having a finite density of sub-gap states?

To address this question, we use a simple description of the sub-gap states in terms of an imaginary contribution to the parent superconductor Green's function that is obtained through the substitution $\omega\longrightarrow \omega+ i\delta$ in $G_{sc}(\omega)$, where
$\delta$ is a  magnetic field-dependent phenomenological parameter that describes the finite density of sub-gap states. We note that developing models for the parent superconductor based on a detailed understanding of the low-energy physics at the microscopic level represents an outstanding problem in this field.\cite{Cole2016,Hui2015} We also note that a finite density of sub-gap states is equivalent to having an additional (i.e., non-superconducting)  equilibrium bath,\cite{Martin2014} which generates dissipation. In turn, the presence of dissipation introduces particle-hole asymmetry into the finite energy  conductance spectrum.\cite{Martin2014,Bauriedl1981,Yazdani1997} The importance of this mechanism in understanding certain  features  of  the  measured conductance  data in semiconductor-superconductor hybrid structures was emphasized recently.\cite{DSarma2016,Liu2017}

\begin{figure}[t]
\begin{center}
\includegraphics[width=0.48\textwidth]{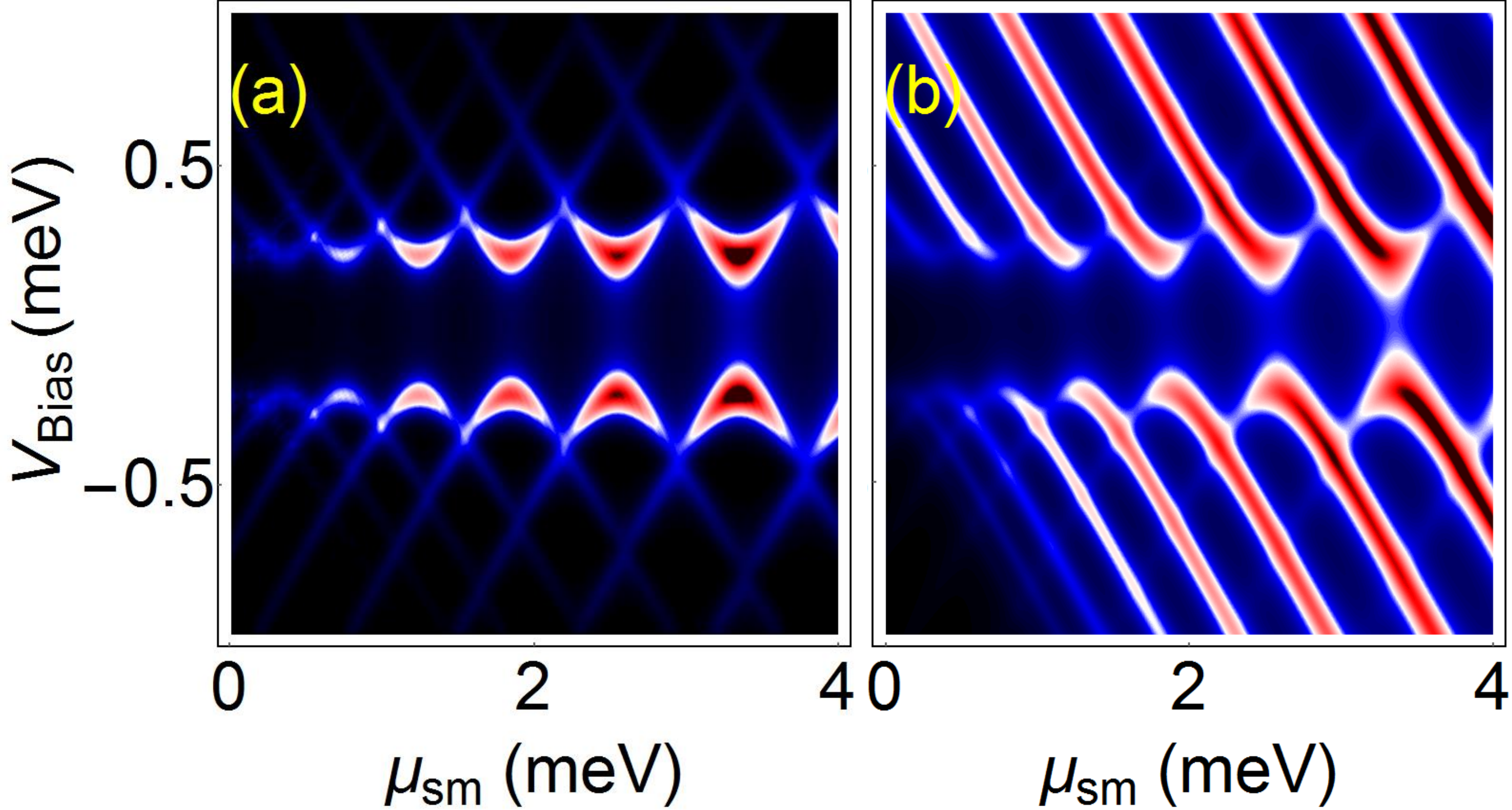}
\end{center}
\vspace{-2mm}

\caption{(Color online) Differential conductance as function of chemical potential and bias voltage for a hybrid system containing (a) a ``clean'' superconductor and (b) a parent superconductor with sub-gap states. The finite density of sub-gap states is modeled as a finite imaginary part in the retarded Green's function of the parent superconductor corresponding to $\delta=20~\mu$eV. The model parameters are:  $L=0.4~\mu$m, $\Delta_0=2~$meV, $\gamma=0.3~$meV, $\Gamma=0$, and $k_BT=2~\mu$eV.}
\label{dis}
\vspace{-3mm}
\end{figure}

Consider now the dependence of the differential conductance of a short wire $L=0.4~\mu m$ on the chemical potential and bias voltage as in Fig. \ref{dis}.  Here, $\Delta_0=2~meV$ and $\gamma=0.3~meV$ As shown in  Fig. \ref{dis} (a), 
in the case of a ``clean'' parent superconductor the features are particle-hole symmetric and strongly suppressed at large bias voltages, as discussed in the context of Fig. \ref{comp}. By contrast, a finite density of sub-gap states in the parent superconductor introduces particle-hole asymmetry, as shown in panel (b) of Fig. \ref{dis}. Also note that the particle-hole asymmetric ``stripy'' features have similar slopes at positive and negative values of $V_{Bias}$ and are not suppressed at large bias voltages.  These types of features are present in the experimentally measured differential conductance of SM-SC hybrid structures\cite{Chen2016}, suggesting that parent superconductors such as NbTiN may have a non-vanishing density of sub-gap states. 

\begin{figure}[t]
\begin{center}
\includegraphics[width=0.47\textwidth]{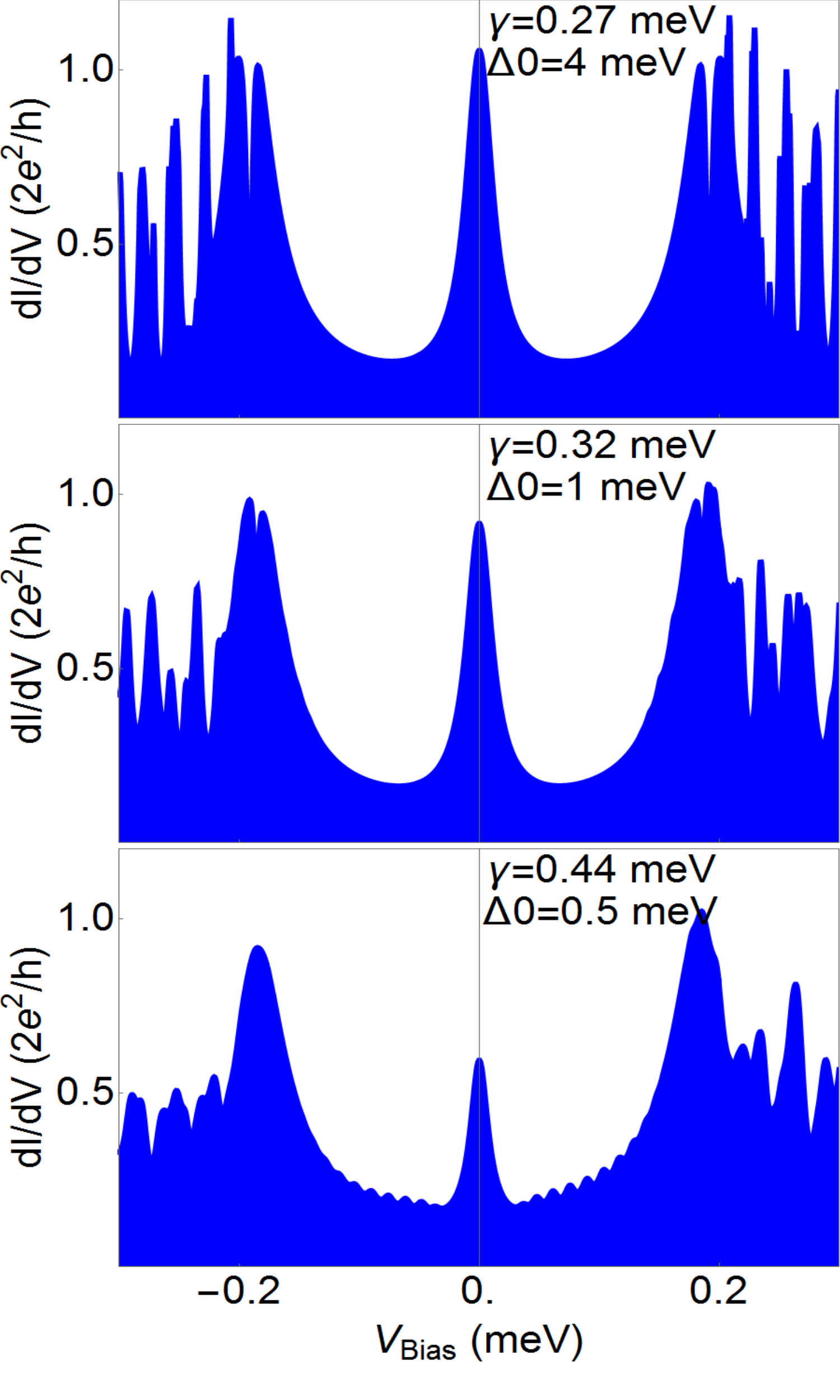}
\end{center}
\vspace{-3.5mm}

\caption{(Color online) Differential conductance for three different values of the ratio between the SM-SC coupling and the parent SC gap.  The induced gap (at zero field), $\Delta_{ind}=0.25~$meV, is the same in all three panels.  The Zeeman field $\Gamma=0.5~$meV is above the critical field for each plot.  The calculation is for a two-band model with a density of subgap states corresponding to $\delta=10~\mu$eV, zero temperature, and a wire length $L=2~\mu$m.  }
\label{new}
\vspace{-3mm}
\end{figure}

We note that these results are consistent with the existence of some generic dissipation\cite{Liu2017}.  Here, we propose the presence of a small but finite density of sub-gap states in the parent superconductor as the natural mechanism responsible for dissipation effects. Thus, dissipation is just one among several important low-energy effects that are directly linked to the parent superconductor, which can be captured theoretically by explicitly incorporating the bulk SC into the modeling. Two relevant examples -- the dependence of the quasiparticle spectral weight within the SM wire on the SM-SC coupling and the renormalization of the low-energy spectrum -- are discussed below. Explicitly incorporating the parent SC into the theory provides a unified framework for describing these effects, which otherwise require different \textit{ad hoc} ingredients.  This suggests that a thorough investigation of the parent superconductor (in the presence of disorder and finite magnetic fields) is absolutely necessary for understanding in detail the low-energy physics of the hybrid structure.

Figure \ref{new} shows differential conductance curves in the topological regime for three different  values of the ratio between the SM-SC coupling and the parent SC gap.  The induced gap is kept constant.  Notice that the height of the ZBP decreases with increasing $\gamma/\Delta_0$.  This dependence is the result of the dissipation effect generated by the non-vanishing density of subgap states in combination with the reduction of the spectral weight within the SM wire associated with the Majorana mode. More specifically, as the (relative) coupling strength $\gamma/\Delta_0$ increases, the spectral weight of the low-energy states (including the Majorana mode) is transfered from the SM wire to the parent SC.\cite{Stanescu2017a} Consequently, for a given value of the tunnel barrier transparency, the weight of the corresponding differential conductance peak decreases as the coupling strength increases, as illustrated in Fig. \ref{new}. The shape of the ZBP will be studied in more detail in the next section. 

\begin{figure}[t]
\begin{center}
\includegraphics[width=0.47\textwidth]{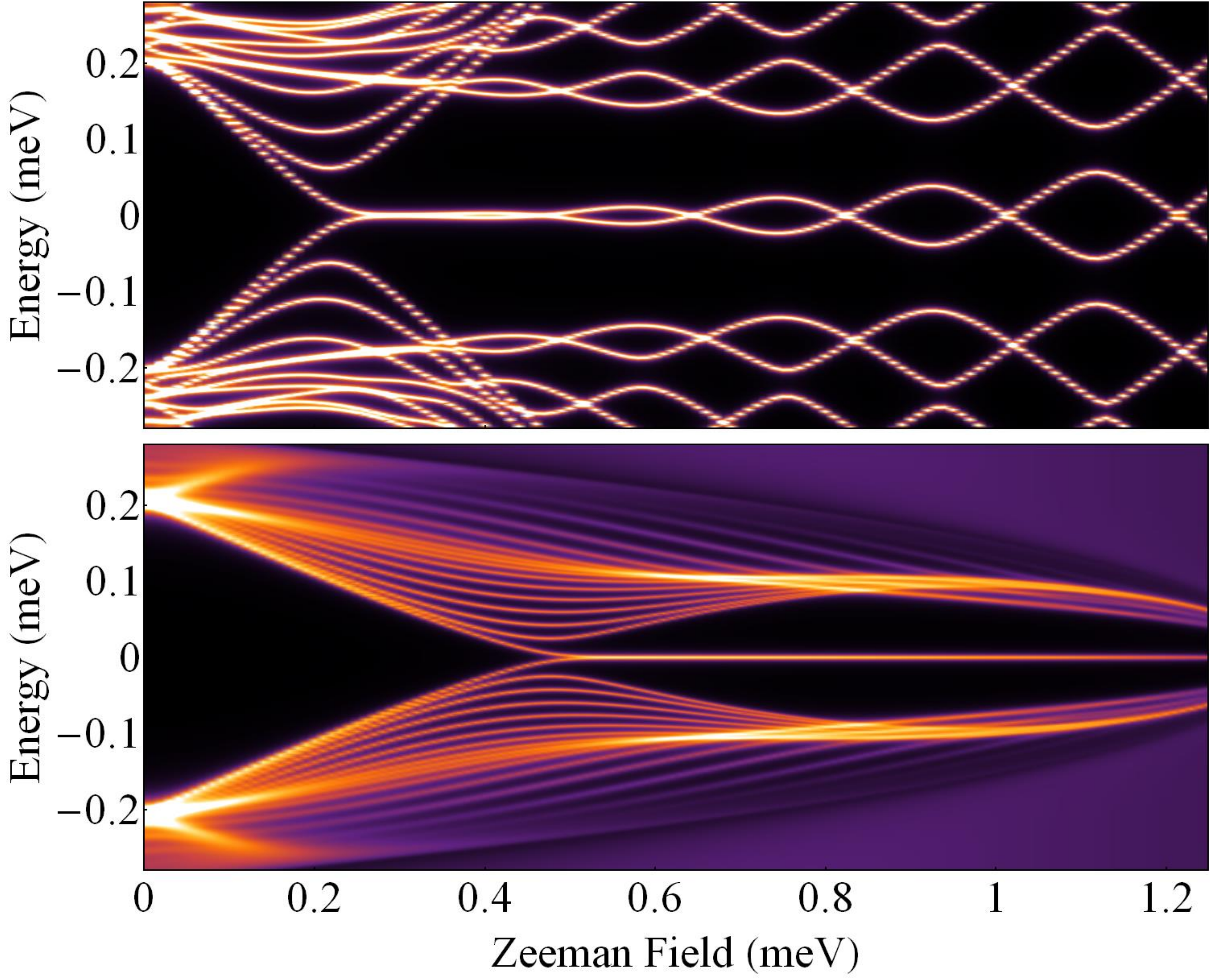}
\end{center}
\vspace{-3.5mm}

\caption{(Color online) Density of states within the SM wire as function of the Zeeman field for a wire with effective pairing potential (model I, top panel) and a 1D system that explicitly includes the parent SC (model II, bottom). The model parameters are given in the main text. Note that all the states in the upper panel have spectral weight equal to 1, while in the lower panel some of the spectral weight is transfered to the parent SC. Also note the strong low-energy renormalization induced by the coupling to the SC, in particular the suppression of the Majorana energy splitting oscillations.} 
\label{SF1}
\vspace{-3mm}
\end{figure}

The dependence of the ZBP height on the coupling strength cannot be captured using an effective pairing potential model with an \textit{ad hoc} dissipation term. To emphasize the importance of explicitly including the parent SC, we compare the dependence of the low-energy spectrum on the applied Zeeman field for i) a 1D wire with an effective pairing potential (model I) and ii) a 1D system that explicitly includes the parent SC (model II). The length of the SM wire is $L=1.5~\mu$m and the SM-SC coupling strength is  $\gamma=0.447~$meV. To better simulate the experimental conditions, we assume that the bulk SC gap $\Delta_0$ is suppressed by the applied magnetic field. Explicitly, we have $\Delta_0(\Gamma) = \widetilde{\Delta}_0\sqrt{1-\Gamma/\Gamma^*}$, with $\widetilde{\Delta}_0=0.3~$meV and $\Gamma^*=1.35~$meV. The corresponding induced gap (defined as the minimum quasiparticle gap at zero field) is\cite{Stanescu2017a} $\Delta_{\rm ind}=0.2~$meV. Since the induced gap is one of the most robust and experimentally accessible features, we consider an effective pairing model characterized by the same value of the induced gap and a field-dependent pairing of the form $\Delta(\Gamma) = \Delta_{\rm ind}\sqrt{1-\Gamma/\Gamma^*}$. Finally, we assume that the parent SC has sub-gap states corresponding to $\delta=12~\mu$eV. A corresponding dissipation term is included in the effective pairing potential model. We expect $\delta$ to be strongly dependent on the applied magnetic field, but we do not incorporate this effect. 

\begin{figure}[t]
\begin{center}
\includegraphics[width=0.47\textwidth]{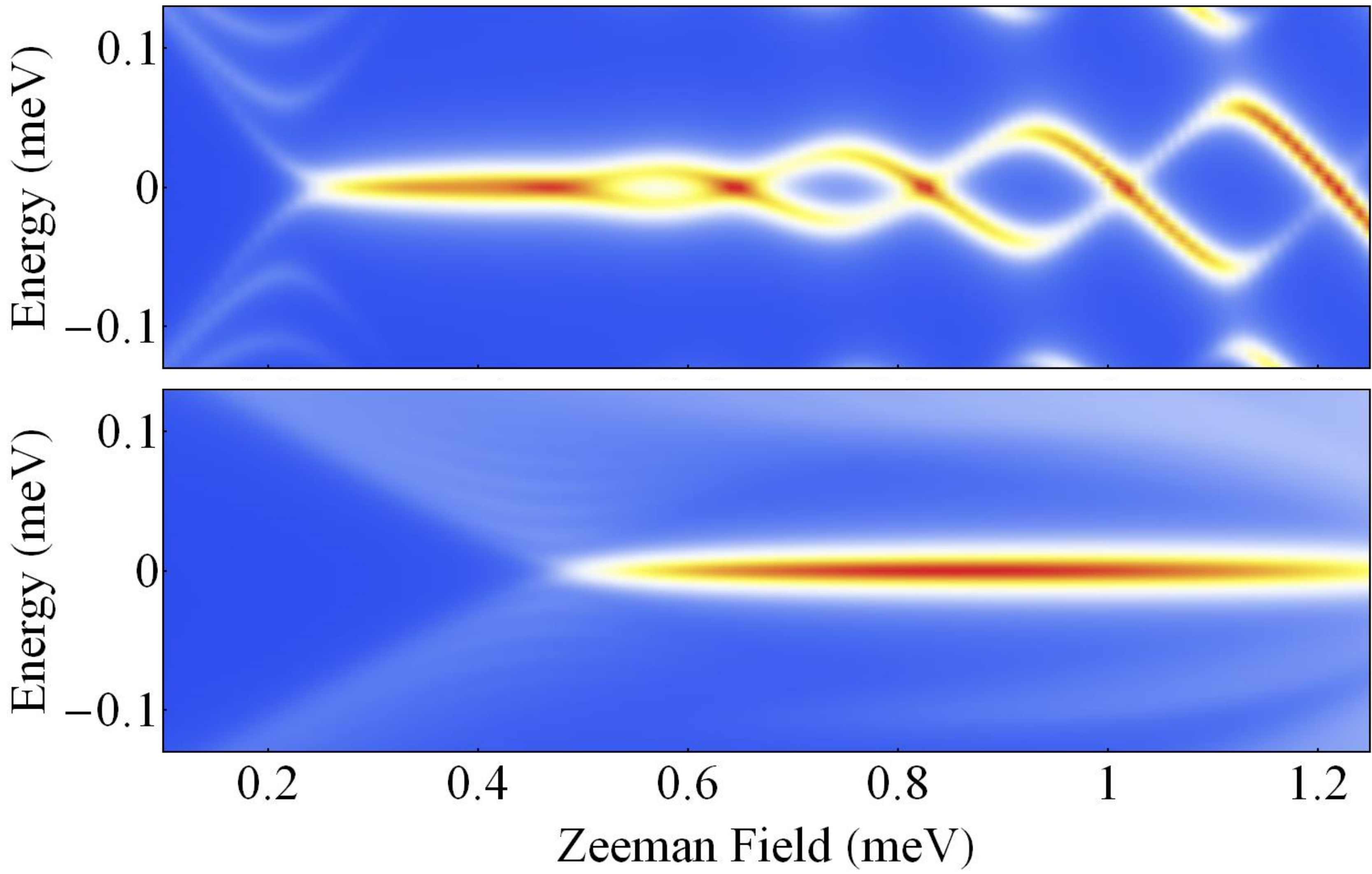}
\end{center}
\vspace{-3.5mm}

\caption{(Color online) Low-energy differential conductance as function of energy (i.e. bias) and Zeeman field for the effective pairing model (top) and the 1D model that incorporates the parent SC (bottom). The corresponding densities of states are shown in Fig. \ref{SF1}.}
\label{SF2}
\vspace{-3mm}
\end{figure}

The dependence of the density of states within the wire on the Zeeman field is shown in Fig. \ref{SF1}, while the signature of the Majorana mode in the differential conductance is illustrated in Fig. \ref{SF2}. Several observations are warranted. First, the two models predict different values of the critical field corresponding to the emergence of the Majorana mode, despite the fact that they are characterized by the same value of the induced gap at $\Gamma=0$. Indeed, assuming that the chemical potential is $\mu=0$, the effective pairing approximation (model I) predicts a critical field given by the induced gap, $\Gamma_c=\Delta_{\rm ind}$, while incorporating the parent SC (model II) gives a critical field controlled by the SM-SC coupling strength,\cite{Stanescu2017a} $\Gamma_c=\gamma$. Of course, one can construct effective pairing models that give  the same value of the critical field as model II, but the corresponding quasiparticle gaps will be significantly overestimated in the intermediate and strong coupling regimes, i.e. for $\gamma > \Delta_0$. Second, we note the strong energy renormalization that leads to the suppression of the Majorana splitting oscillations in model II. One can still have Majorana splitting in the intermediate/strong coupling regime, but it requires very short wires. Note that in model I the collapse of the SC gap at $\Gamma^*$ has no visible impact on the amplitude of the oscillations, which is increasing with the Zeeman field (see top panels in Figs. \ref{SF1} and \ref{SF2}). On the other hand, in model II the collapse of the SC gap results in a reduction of the ZBP height. This is another manifestation of the dependence of the quasiparticle spectral weight on $\gamma/\Delta_0$. Since $\Delta_0$ is field-dependent, the relative SM-SC coupling strength increases with $\Gamma$, which results in a decrease of the spectral weight of the Majorana mode reflected in the conductance. We emphasize that in these calculations the chemical potential is constant, hence the Fermi \textit{k}-vector and the corresponding Fermi velocity of the Majorana band increase with the applied field. Consequently, the transparency of the barrier increases with $\Gamma$.  In model I, this results in a ZBP that becomes stronger with $\Gamma$; i.e., its weight increases with the field and so does its height (if it is below the quantized value at finite $\delta$ and/or $T$) whenever the wire is long-enough (to avoid energy splitting).  In model II, on the other hand, the effect of a higher barrier transparency is offset by the loss of spectral weight, the net result being the eventual  decrease of the ZBP strength upon approaching $\Gamma^*$. 
Finally, we note that the dissipation-induced particle-hole asymmetry responsible for the ``stripy'' features in Fig. \ref{dis}  is present even at very low energies, as illustrated in the top panel of Fig. \ref{SF2}. This effect could explain recent experimental observations\cite{Chen2016} showing similar low-energy features. 

\subsection{The shape of the zero bias conductance peak} \label{S_IIID}

\begin{figure}[t!]
\begin{center}
\includegraphics[width=0.48\textwidth]{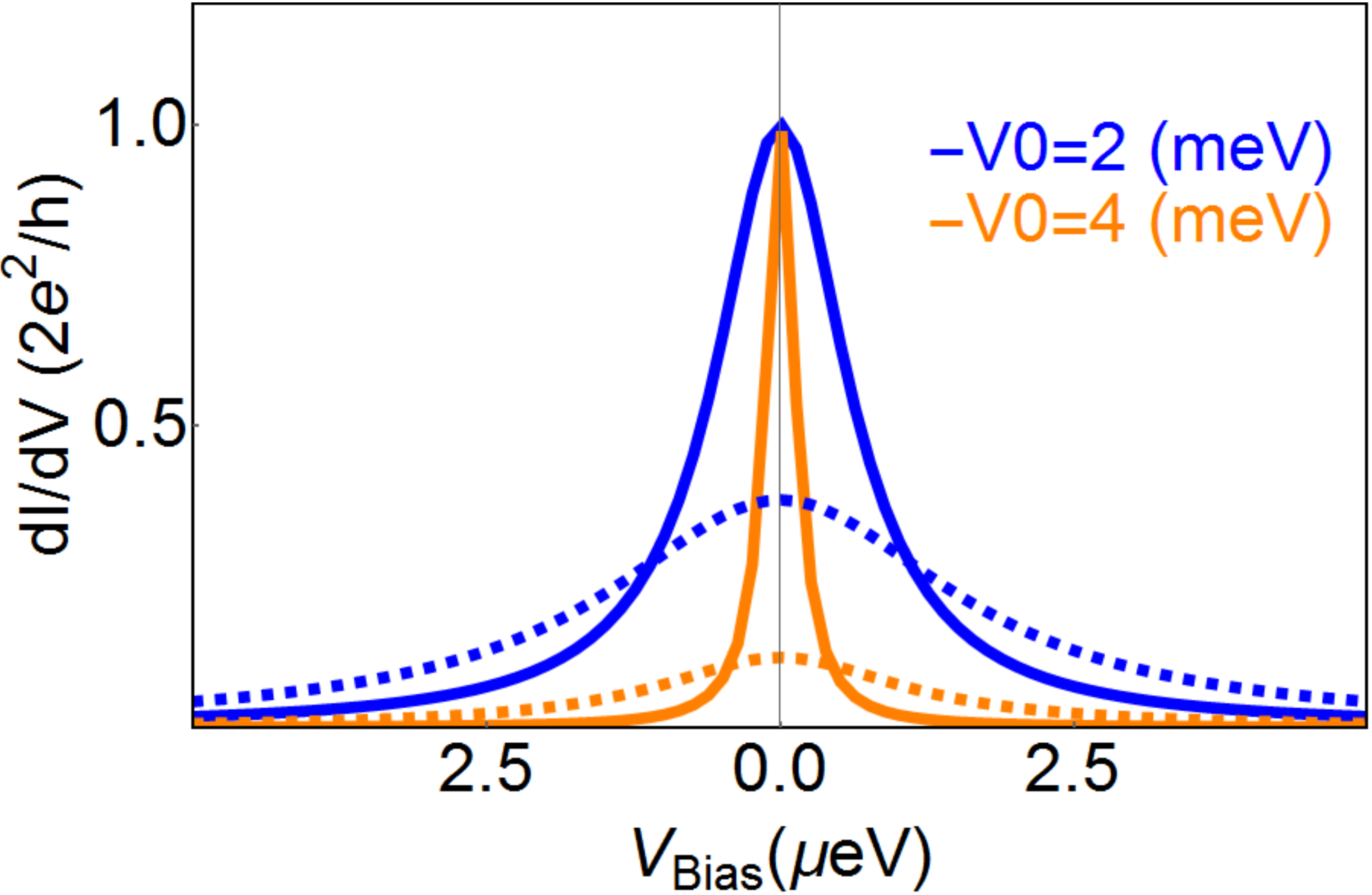}
\end{center}
\vspace{-2mm}
\caption{(Color online) Majorana-induced zero-bias conductance peaks for two different values of the potential barrier. The solid lines correspond to a clean system, while the dashed lines are for a parent superconductor with a finite density of sub-gap states corresponding to $\delta=20~\mu$eV. Note that the quantization of the zero-bias peak is broken in the presence of a finite density of sub-gap states in the parent superconductor.  The calculation was done for a nanowire of length $L=0.2~\mu$m, SM-SC coupling $\gamma=0.25~$meV, and parent SC gap $\Delta_0=2~$meV.  The Zeeman field is $\Gamma=0.3~$meV.}
\label{ZBP}
\end{figure}

We conclude our analysis with a summary of the dependence of the Majorana-induced zero-bias conductance peak on the relevant system parameters. In this section we will study the ZBP of a hybrid system consisting of a very short nanowire of length $L=0.2~\mu$m coupled to a  superconductor with $\Delta_0=2~$meV, the effective coupling strength being $\gamma=0.25~$meV.  The Zeeman field is set to $\Gamma=0.3~$meV.  We note that having a short wire ensures that the Majorana mode is separated by a large gap from finite energy excitations, so that the area associated with the zero-bias conduction peak can be accurately estimated.  
Figure \ref{ZBP} shows the Majorana-induced zero-bias peak (ZBP) at $T=0$ for two different values of the potential barrier in a system with either a clean parent superconductor (solid lines) or a superconductor with a finite density of sub-gap states (dashed lines). First, we note that in the clean system the ZBPs are quantized at $2e^2/h$ and that their widths (and, implicitly, the areas under the corresponding $dI/dV$ curves) are reduced by increasing the barrier potential (i.e., reducing the transparency of the barrier).  Second, in the structure with a ``dirty'' parent superconductor (dashed lines in Fig. \ref{ZBP}) the height of the ZBP is significantly lower than the quantized value. In fact,  the effect of sub-gap states being present in the superconductor is to introduce a broadening of the Majorana mode. The area under the  $dI/dV$ curve is the same as in the clean system (for a given potential barrier), but the width of the peak is intrinsically finite. In other words, increasing the barrier height generates a thin quantized peak in the clean system, but results in a small wide peak in the presence of sub-gap states (see solid and dashed orange/light gray lines in Fig. \ref{ZBP}). Intuitively, we can understand the broadening of the ZBP as a result of the Majorana mode hybridizing with the sub-gap states from the parent superconductor. The resulting hybrid states will have energies distributed within a certain $\delta$-dependent energy window about zero. As a result, the LDOS at the end of the wire will be characterized by a finite width peak centered at $E=0$.\cite{Stanescu2017a} By contrast, in a clean system the corresponding signature of the Majorana mode is a delta function-type contribution to the LDOS. Note that the broadening effect of the sub-gap states is qualitatively similar to the effect of finite temperature.

\begin{figure}[t!]
\begin{center}
\includegraphics[width=0.4\textwidth]{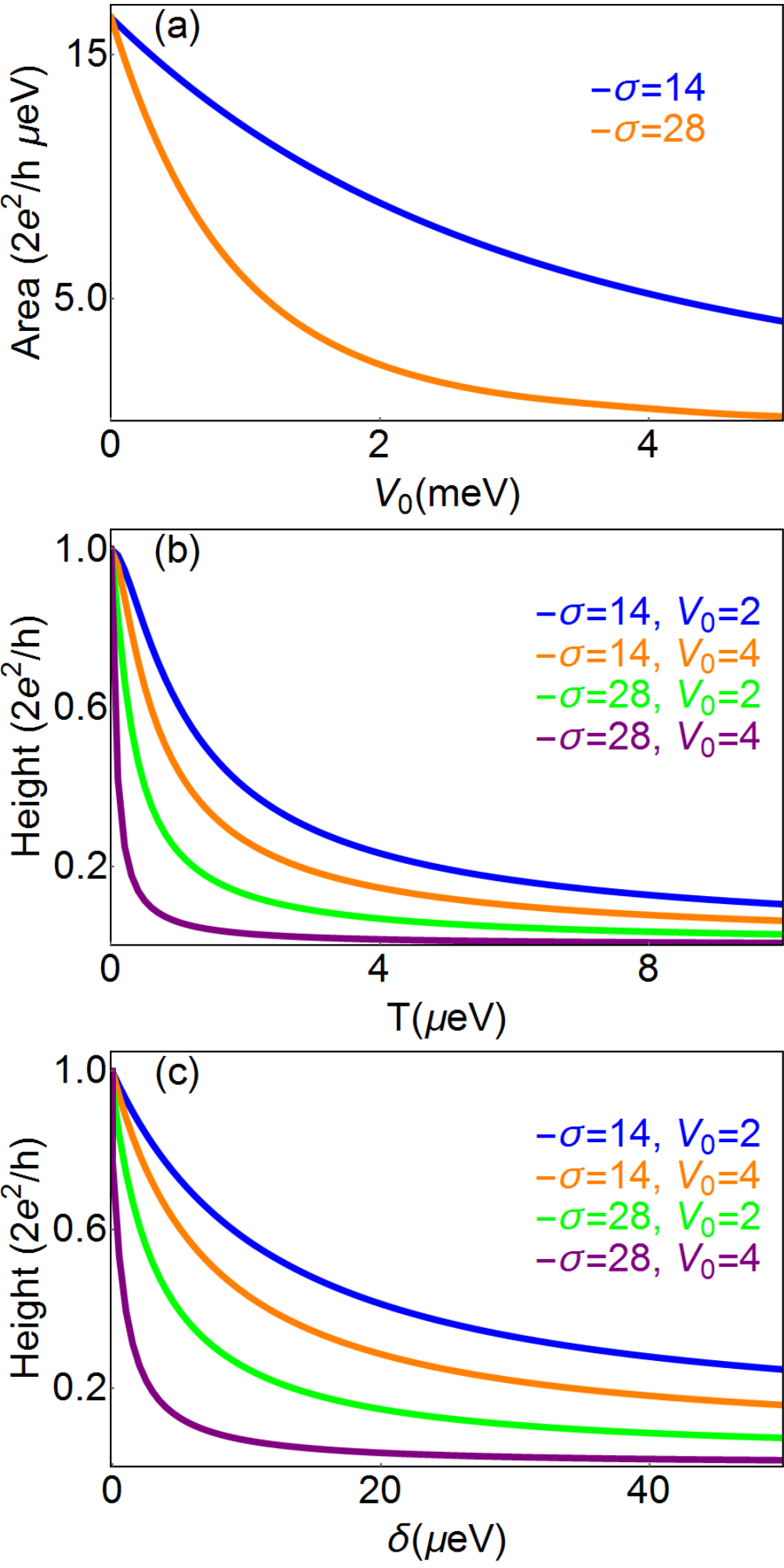}
\end{center}
\vspace{-2mm}

\caption{(Color online) Dependence of the zero-bias conduction peak on relevant parameters. (a) The area of the ZBP as a function of the potential barrier height for two different barrier widths.  The barrier potential is modeled as a Gaussian with a standard deviation $\sigma$.  Note that for $T=0$, $\delta=0$ the peaks are quantized (at $2e^2/h$). (b) The height of the ZBP as a function of temperature for $\delta=0$ and various widths and heights of the tunnel barrier potential. Note that the area of the ZBP is independent of $T$. (c) The height  of the ZBP as a function of $\delta$ for $T=0$  and various widths and heights of the barrier potential. The ZBP area is $\delta$-independent.  The calculation was done for a nanowire of length $L=0.2~\mu$m, SM-SC coupling $\gamma=0.25~$meV, and parent SC gap $\Delta_0=2~$meV.  The Zeeman field is $\Gamma=0.3~$meV.}
\label{ZBP0}
\end{figure}

The dependence of the zero-bias conduction peak on the transparency of the barrier, temperature, and density of sub-gap states is shown in Fig. \ref{ZBP0}. First, we note that the transparency of the barrier depends on both the height and width of the barrier potential. For simplicity, we model the potential barrier as a Gaussian, $V_b(x) = V_0\exp[-(x-x_0)^2/2\sigma^2]$. Hence,  the height of the barrier is parametrized by $V_0$ and its width by the standard deviation $\sigma$. Increasing $V_0$ or $\sigma$ reduces the transparency of the barrier, which results in a smaller area under the $dI/dV$ curve [see panel (a)]. Note that in a clean system at $T=0$ the height of the ZBP is quantized for all values of $V_0$ and $\sigma$. For a given tunnel barrier potential, introducing finite temperature or finite density of sub-gap states (in the parent SC) does not modify the area of the ZBP, but broadens the peak. Consequently, its height gets reduced. The dependence of the ZBP height on $T$ and $\delta$ for different potential barriers is shown in panels (b) and (c), respectively. Note the similarities between the dependence of the  ZBP height on temperature\cite{Lin2012,Liu2017} (panel b) and on the density of subgap states (panel c). 

In the high temperature limit, which is barrier-dependent, see Fig. \ref{ZBP0} (b), the dependence of the height of the ZBP on temperature is given by the analytical expression $A/(8T)$, where $A$ is the area of the ZBP.  Similarly, in the limit of high density of sub-gap states, the dependence of the height of the ZBP on $\delta$ is given by $A \pi/(2\delta)$.  As the zero-bias peak approaches its quantized value with lowering $T$ (or $\delta$), the dependence of its height on temperature (or on the density of sub-gap states) becomes more complicated. We note that all the curves in panel (b) can be collapsed into a single curve corresponding to a ZBP of area $A_0$ through the scaling transformation  $T\rightarrow\frac{A_0}{A}T$, where $A$ is the area under the ZBP corresponding to an arbitrary curve. 
  Similarly, any two curves  in panel (c) are related to one another by the transformation $(\delta\rightarrow\frac{A_2}{A_1}\delta)$ and can be collapsed into a single curve.  However, the transformation $(T\rightarrow 4\pi\delta)$ mapping a curve with area $A$ in panel (b) to a curve in panel (c) with the same area under the ZBP only works in the high-temperature/high density of sub-gap states regimes.

We note that {\em all} transport experiments on semiconductor-superconductor hybrid structures reported so far have observed zero-bias peaks having heights significantly smaller than the (theoretically predicted) quantized value, $2e^2/h$. These small values are inconsistent with expectations based on the nominal temperature of the system. Of course, the effective temperature could be much higher. However, the presence of sub-gap states is a natural alternative explanation for the low values of the observed zero-bias conduction peaks. 

\section{Conclusions} \label{S_IV}

We have calculated the tunneling conductance of a normal metal - semiconductor wire - superconductor structure within the Blonder-Tinkham-Klapwijk and Keldysh formalisms by treating the parent superconductor as an active component of the hybrid system. This treatment ensures that several key effects that  control the low-energy physics of the hybrid structure are accurately captured by the theory. These effects include the proximity-induced renormalization  of the low-energy states, including the renormalization of the topological gap that protects the Majorana zero modes,\cite{Stanescu2017a} the emergence of additional conductance peaks at  energies corresponding to the gap edge  of the parent superconductor\cite{Cole2015,Reeg2017,Stanescu2017a}, and the broadening of the low-energy features due to hybridization of the states within the semiconductor wire with sub-gap states  in the parent superconductor\cite{Stanescu2017a}. 

We  find that the differential conductance is a good indicator for the local density of states at the end on the hybrid structure adjacent to the tunnel barrier. This includes both the semiconductor wire and the nearby parent superconductor. A clearly defined feature associated with the parent superconductor emerges at energies corresponding to the bulk gap.  In the intermediate/strong coupling regime the features associated with sub-gap states (e.g., the induced gap) and those corresponding to contributions from the parent superconductor have similar characteristic energies. We predict that one can disentangle them by varying the strength of the potential barrier, which has a stronger effect on the sub-gap features. For example, in the high barrier limit one can almost completely suppress the sub-gap features while having a measurable parent superconductor contribution. In the topological regime, varying the transparency of the tunnel barrier changes the area of the zero-bias conductance peak, while its height is quantized ($2e^2/h$)  at zero temperature. Finite temperature reduces the height of the ZBP but preserves its area.

The most dramatic implication of treating the parent superconductor as an active component of the system occurs if we assume that the superconductor itself has sub-gap states induced by disorder and finite magnetic fields. These states hybridize with the low-energy states in the semiconductor wire significantly altering their properties. We find that in the presence of sub-gap states in the parent superconductor the differential conductance develops characteristic ``stripy'' features (as function of the chemical potential and bias voltage) that break particle-hole symmetry. In addition, the presence of sub-gap states has an effect similar to finite temperature and destroys the quantization of the Majorana-induced zero-bias conductance peak. More generally, the presence of these states results in a broadening of the low-energy features that characterize the hybrid structure.
From the perspective of quantum computation,   this is a deadly effect, as it destroys the topological gap that protects the low-energy Majorana subspace. All the signatures identified here as being associated with the presence of sub-gap states in the parent superconductor can be recognized in the transport measurements on semiconductor-superconductor hybrid structures reported so far. Therefore, we find it imperative to understand in detail the low-energy properties of currently-used parent superconductors, such as Al and NbTiN, and to consider a systematic effort to optimize their properties or to engineer better suited parent superconductors. 

In summary, we have shown that explicitly treating the parent superconductor as an active component of the semiconductor-superconductor structure generates a number of features in the calculated differential tunneling conductance that would otherwise be absent. Some of these features have not been previously discussed in any detail. One of the new features is the  reduction in weight of the differential conductance peaks  as the coupling to the parent superconductor increases, as shown in Fig. \ref{D0}. Since the reduction (and eventual collapse) of the parent SC gap at high magnetic fields implies an increase of the effective coupling (relative to the SC gap),  a consequence of this feature is the decrease of the ZBP strength at high Zeeman fields, as illustrated in Fig. \ref{SF2}. Another feature is the presence of a superconducting resonance at the energy scale corresponding to the bulk gap (see Fig. \ref{scnsc}) with an amplitude that has a dependence on the tunnel barrier height different from that of sub-gap (induced) peaks, as shown in  Fig. \ref{V0}. In addition, the possible presence of sub-gap states in the parent superconductor results in particle-hole asymmetric ``stripy'' features in the differential conductance, like those illustrated in Fig.  \ref{dis}, that persist even at very low energy, as shown in  Fig. \ref{SF2}. Finally, explicitly considering the parent superconductor can strongly suppress the Majorana energy splitting oscillations and generate a finite density of states inside the topological gap, as shown in Fig. \ref{SF2}. Considering that all these features, some of them new, which were calculated within a unitary framework that explicitly incorporates the superconductor, have been observed experimentally in tunneling measurements on semiconductor-superconductor structures, we conclude that our work provides strong evidence regarding the critical role of the parent superconductor in Majorana devices.

~

This work is supported by NSF DMR-1414683.				
												
\bibliography{REFERENCES}

\begin{thebibliography}{54}%
\makeatletter
\providecommand \@ifxundefined [1]{%
 \@ifx{#1\undefined}
}%
\providecommand \@ifnum [1]{%
 \ifnum #1\expandafter \@firstoftwo
 \else \expandafter \@secondoftwo
 \fi
}%
\providecommand \@ifx [1]{%
 \ifx #1\expandafter \@firstoftwo
 \else \expandafter \@secondoftwo
 \fi
}%
\providecommand \natexlab [1]{#1}%
\providecommand \enquote  [1]{``#1''}%
\providecommand \bibnamefont  [1]{#1}%
\providecommand \bibfnamefont [1]{#1}%
\providecommand \citenamefont [1]{#1}%
\providecommand \href@noop [0]{\@secondoftwo}%
\providecommand \href [0]{\begingroup \@sanitize@url \@href}%
\providecommand \@href[1]{\@@startlink{#1}\@@href}%
\providecommand \@@href[1]{\endgroup#1\@@endlink}%
\providecommand \@sanitize@url [0]{\catcode `\\12\catcode `\$12\catcode
  `\&12\catcode `\#12\catcode `\^12\catcode `\_12\catcode `\%12\relax}%
\providecommand \@@startlink[1]{}%
\providecommand \@@endlink[0]{}%
\providecommand \url  [0]{\begingroup\@sanitize@url \@url }%
\providecommand \@url [1]{\endgroup\@href {#1}{\urlprefix }}%
\providecommand \urlprefix  [0]{URL }%
\providecommand \Eprint [0]{\href }%
\providecommand \doibase [0]{http://dx.doi.org/}%
\providecommand \selectlanguage [0]{\@gobble}%
\providecommand \bibinfo  [0]{\@secondoftwo}%
\providecommand \bibfield  [0]{\@secondoftwo}%
\providecommand \translation [1]{[#1]}%
\providecommand \BibitemOpen [0]{}%
\providecommand \bibitemStop [0]{}%
\providecommand \bibitemNoStop [0]{.\EOS\space}%
\providecommand \EOS [0]{\spacefactor3000\relax}%
\providecommand \BibitemShut  [1]{\csname bibitem#1\endcsname}%
\let\auto@bib@innerbib\@empty
\bibitem [{\citenamefont {Hasan}\ and\ \citenamefont {Kane}(2010)}]{Hasan2010}%
  \BibitemOpen
  \bibfield  {author} {\bibinfo {author} {\bibfnamefont {M.~Z.}\ \bibnamefont
  {Hasan}}\ and\ \bibinfo {author} {\bibfnamefont {C.~L.}\ \bibnamefont
  {Kane}},\ }\href {\doibase 10.1103/RevModPhys.82.3045} {\bibfield  {journal}
  {\bibinfo  {journal} {Rev. Mod. Phys.}\ }\textbf {\bibinfo {volume} {82}},\
  \bibinfo {pages} {3045} (\bibinfo {year} {2010})}\BibitemShut {NoStop}%
\bibitem [{\citenamefont {Qi}\ and\ \citenamefont {Zhang}(2011)}]{Qi2011}%
  \BibitemOpen
  \bibfield  {author} {\bibinfo {author} {\bibfnamefont {X.-L.}\ \bibnamefont
  {Qi}}\ and\ \bibinfo {author} {\bibfnamefont {S.-C.}\ \bibnamefont {Zhang}},\
  }\href {\doibase 10.1103/RevModPhys.83.1057} {\bibfield  {journal} {\bibinfo
  {journal} {Rev. Mod. Phys.}\ }\textbf {\bibinfo {volume} {83}},\ \bibinfo
  {pages} {1057} (\bibinfo {year} {2011})}\BibitemShut {NoStop}%
\bibitem [{\citenamefont {Bernevig}(2013)}]{Bernevig2013}%
  \BibitemOpen
  \bibfield  {author} {\bibinfo {author} {\bibfnamefont {B.~A.}\ \bibnamefont
  {Bernevig}},\ }\href@noop {} {\emph {\bibinfo {title} {{Topological
  Insulators and Topological Superconductors}}}}\ (\bibinfo  {publisher}
  {Princeton University Press, New Jersey},\ \bibinfo {year}
  {2013})\BibitemShut {NoStop}%
\bibitem [{\citenamefont {Alicea}(2012)}]{Alicea2012}%
  \BibitemOpen
  \bibfield  {author} {\bibinfo {author} {\bibfnamefont {J.}~\bibnamefont
  {Alicea}},\ }\href {http://stacks.iop.org/0034-4885/75/i=7/a=076501}
  {\bibfield  {journal} {\bibinfo  {journal} {Reports on Progress in Physics}\
  }\textbf {\bibinfo {volume} {75}},\ \bibinfo {pages} {076501} (\bibinfo
  {year} {2012})}\BibitemShut {NoStop}%
\bibitem [{\citenamefont {Leijnse}\ and\ \citenamefont
  {Flensberg}(2012)}]{Leijnse2012}%
  \BibitemOpen
  \bibfield  {author} {\bibinfo {author} {\bibfnamefont {M.}~\bibnamefont
  {Leijnse}}\ and\ \bibinfo {author} {\bibfnamefont {K.}~\bibnamefont
  {Flensberg}},\ }\href {http://stacks.iop.org/0268-1242/27/i=12/a=124003}
  {\bibfield  {journal} {\bibinfo  {journal} {Semiconductor Science and
  Technology}\ }\textbf {\bibinfo {volume} {27}},\ \bibinfo {pages} {124003}
  (\bibinfo {year} {2012})}\BibitemShut {NoStop}%
\bibitem [{\citenamefont {Beenakker}(2013)}]{Beenakker2013}%
  \BibitemOpen
  \bibfield  {author} {\bibinfo {author} {\bibfnamefont {C.}~\bibnamefont
  {Beenakker}},\ }\href {\doibase 10.1146/annurev-conmatphys-030212-184337}
  {\bibfield  {journal} {\bibinfo  {journal} {Annual Review of Condensed Matter
  Physics}\ }\textbf {\bibinfo {volume} {4}},\ \bibinfo {pages} {113} (\bibinfo
  {year} {2013})},\ \Eprint
  {http://arxiv.org/abs/http://dx.doi.org/10.1146/annurev-conmatphys-030212-184337}
  {http://dx.doi.org/10.1146/annurev-conmatphys-030212-184337} \BibitemShut
  {NoStop}%
\bibitem [{\citenamefont {Stanescu}\ and\ \citenamefont
  {Tewari}(2013)}]{Stanescu2013}%
  \BibitemOpen
  \bibfield  {author} {\bibinfo {author} {\bibfnamefont {T.~D.}\ \bibnamefont
  {Stanescu}}\ and\ \bibinfo {author} {\bibfnamefont {S.}~\bibnamefont
  {Tewari}},\ }\href {http://stacks.iop.org/0953-8984/25/i=23/a=233201}
  {\bibfield  {journal} {\bibinfo  {journal} {J. Phys.: Condens. Matter}\
  }\textbf {\bibinfo {volume} {25}},\ \bibinfo {pages} {233201} (\bibinfo
  {year} {2013})}\BibitemShut {NoStop}%
\bibitem [{\citenamefont {Nayak}\ \emph {et~al.}(2008)\citenamefont {Nayak},
  \citenamefont {Simon}, \citenamefont {Stern}, \citenamefont {Freedman},\ and\
  \citenamefont {Das~Sarma}}]{Nayak2008}%
  \BibitemOpen
  \bibfield  {author} {\bibinfo {author} {\bibfnamefont {C.}~\bibnamefont
  {Nayak}}, \bibinfo {author} {\bibfnamefont {S.~H.}\ \bibnamefont {Simon}},
  \bibinfo {author} {\bibfnamefont {A.}~\bibnamefont {Stern}}, \bibinfo
  {author} {\bibfnamefont {M.}~\bibnamefont {Freedman}}, \ and\ \bibinfo
  {author} {\bibfnamefont {S.}~\bibnamefont {Das~Sarma}},\ }\href {\doibase
  10.1103/RevModPhys.80.1083} {\bibfield  {journal} {\bibinfo  {journal} {Rev.
  Mod. Phys.}\ }\textbf {\bibinfo {volume} {80}},\ \bibinfo {pages} {1083}
  (\bibinfo {year} {2008})}\BibitemShut {NoStop}%
\bibitem [{\citenamefont {Stanescu}(2017)}]{Stanescu2017}%
  \BibitemOpen
  \bibfield  {author} {\bibinfo {author} {\bibfnamefont {T.~D.}\ \bibnamefont
  {Stanescu}},\ }\href@noop {} {\emph {\bibinfo {title} {Introduction to
  topological quantum matter and quantum computation}}}\ (\bibinfo  {publisher}
  {CRC Press, Taylor \& Francis Group},\ \bibinfo {year} {2017})\BibitemShut
  {NoStop}%
\bibitem [{\citenamefont {Lutchyn}\ \emph {et~al.}(2010)\citenamefont
  {Lutchyn}, \citenamefont {Sau},\ and\ \citenamefont
  {Das~Sarma}}]{Lutchyn2010}%
  \BibitemOpen
  \bibfield  {author} {\bibinfo {author} {\bibfnamefont {R.~M.}\ \bibnamefont
  {Lutchyn}}, \bibinfo {author} {\bibfnamefont {J.~D.}\ \bibnamefont {Sau}}, \
  and\ \bibinfo {author} {\bibfnamefont {S.}~\bibnamefont {Das~Sarma}},\ }\href
  {\doibase 10.1103/PhysRevLett.105.077001} {\bibfield  {journal} {\bibinfo
  {journal} {Phys. Rev. Lett.}\ }\textbf {\bibinfo {volume} {105}},\ \bibinfo
  {pages} {077001} (\bibinfo {year} {2010})}\BibitemShut {NoStop}%
\bibitem [{\citenamefont {Oreg}\ \emph {et~al.}(2010)\citenamefont {Oreg},
  \citenamefont {Refael},\ and\ \citenamefont {von Oppen}}]{Oreg2010}%
  \BibitemOpen
  \bibfield  {author} {\bibinfo {author} {\bibfnamefont {Y.}~\bibnamefont
  {Oreg}}, \bibinfo {author} {\bibfnamefont {G.}~\bibnamefont {Refael}}, \ and\
  \bibinfo {author} {\bibfnamefont {F.}~\bibnamefont {von Oppen}},\ }\href
  {\doibase 10.1103/PhysRevLett.105.177002} {\bibfield  {journal} {\bibinfo
  {journal} {Phys. Rev. Lett.}\ }\textbf {\bibinfo {volume} {105}},\ \bibinfo
  {pages} {177002} (\bibinfo {year} {2010})}\BibitemShut {NoStop}%
\bibitem [{\citenamefont {Read}\ and\ \citenamefont {Green}(2000)}]{Read2000}%
  \BibitemOpen
  \bibfield  {author} {\bibinfo {author} {\bibfnamefont {N.}~\bibnamefont
  {Read}}\ and\ \bibinfo {author} {\bibfnamefont {D.}~\bibnamefont {Green}},\
  }\href {\doibase 10.1103/PhysRevB.61.10267} {\bibfield  {journal} {\bibinfo
  {journal} {Phys. Rev. B}\ }\textbf {\bibinfo {volume} {61}},\ \bibinfo
  {pages} {10267} (\bibinfo {year} {2000})}\BibitemShut {NoStop}%
\bibitem [{\citenamefont {Sau}\ \emph {et~al.}(2010)\citenamefont {Sau},
  \citenamefont {Tewari}, \citenamefont {Lutchyn}, \citenamefont {Stanescu},\
  and\ \citenamefont {Das~Sarma}}]{Sau2010}%
  \BibitemOpen
  \bibfield  {author} {\bibinfo {author} {\bibfnamefont {J.~D.}\ \bibnamefont
  {Sau}}, \bibinfo {author} {\bibfnamefont {S.}~\bibnamefont {Tewari}},
  \bibinfo {author} {\bibfnamefont {R.~M.}\ \bibnamefont {Lutchyn}}, \bibinfo
  {author} {\bibfnamefont {T.~D.}\ \bibnamefont {Stanescu}}, \ and\ \bibinfo
  {author} {\bibfnamefont {S.}~\bibnamefont {Das~Sarma}},\ }\href {\doibase
  10.1103/PhysRevB.82.214509} {\bibfield  {journal} {\bibinfo  {journal} {Phys.
  Rev. B}\ }\textbf {\bibinfo {volume} {82}},\ \bibinfo {pages} {214509}
  (\bibinfo {year} {2010})}\BibitemShut {NoStop}%
\bibitem [{\citenamefont {Sengupta}\ \emph {et~al.}(2001)\citenamefont
  {Sengupta}, \citenamefont {\ifmmode \check{Z}\else
  \v{Z}\fi{}uti\ifmmode~\acute{c}\else \'{c}\fi{}}, \citenamefont {Kwon},
  \citenamefont {Yakovenko},\ and\ \citenamefont {Das~Sarma}}]{Sengupta2001}%
  \BibitemOpen
  \bibfield  {author} {\bibinfo {author} {\bibfnamefont {K.}~\bibnamefont
  {Sengupta}}, \bibinfo {author} {\bibfnamefont {I.}~\bibnamefont {\ifmmode
  \check{Z}\else \v{Z}\fi{}uti\ifmmode~\acute{c}\else \'{c}\fi{}}}, \bibinfo
  {author} {\bibfnamefont {H.-J.}\ \bibnamefont {Kwon}}, \bibinfo {author}
  {\bibfnamefont {V.~M.}\ \bibnamefont {Yakovenko}}, \ and\ \bibinfo {author}
  {\bibfnamefont {S.}~\bibnamefont {Das~Sarma}},\ }\href {\doibase
  10.1103/PhysRevB.63.144531} {\bibfield  {journal} {\bibinfo  {journal} {Phys.
  Rev. B}\ }\textbf {\bibinfo {volume} {63}},\ \bibinfo {pages} {144531}
  (\bibinfo {year} {2001})}\BibitemShut {NoStop}%
\bibitem [{\citenamefont {Law}\ \emph {et~al.}(2009)\citenamefont {Law},
  \citenamefont {Lee},\ and\ \citenamefont {Ng}}]{Law2009}%
  \BibitemOpen
  \bibfield  {author} {\bibinfo {author} {\bibfnamefont {K.~T.}\ \bibnamefont
  {Law}}, \bibinfo {author} {\bibfnamefont {P.~A.}\ \bibnamefont {Lee}}, \ and\
  \bibinfo {author} {\bibfnamefont {T.~K.}\ \bibnamefont {Ng}},\ }\href
  {\doibase 10.1103/PhysRevLett.103.237001} {\bibfield  {journal} {\bibinfo
  {journal} {Phys. Rev. Lett.}\ }\textbf {\bibinfo {volume} {103}},\ \bibinfo
  {pages} {237001} (\bibinfo {year} {2009})}\BibitemShut {NoStop}%
\bibitem [{\citenamefont {Flensberg}(2010)}]{Flensberg2010}%
  \BibitemOpen
  \bibfield  {author} {\bibinfo {author} {\bibfnamefont {K.}~\bibnamefont
  {Flensberg}},\ }\href {\doibase 10.1103/PhysRevB.82.180516} {\bibfield
  {journal} {\bibinfo  {journal} {Phys. Rev. B}\ }\textbf {\bibinfo {volume}
  {82}},\ \bibinfo {pages} {180516} (\bibinfo {year} {2010})}\BibitemShut
  {NoStop}%
\bibitem [{\citenamefont {Wimmer}\ \emph {et~al.}(2011)\citenamefont {Wimmer},
  \citenamefont {Akhmerov}, \citenamefont {Dahlhaus},\ and\ \citenamefont
  {Beenakker}}]{Wimmer2011}%
  \BibitemOpen
  \bibfield  {author} {\bibinfo {author} {\bibfnamefont {M.}~\bibnamefont
  {Wimmer}}, \bibinfo {author} {\bibfnamefont {A.~R.}\ \bibnamefont
  {Akhmerov}}, \bibinfo {author} {\bibfnamefont {J.~P.}\ \bibnamefont
  {Dahlhaus}}, \ and\ \bibinfo {author} {\bibfnamefont {C.~W.~J.}\ \bibnamefont
  {Beenakker}},\ }\href {http://stacks.iop.org/1367-2630/13/i=5/a=053016}
  {\bibfield  {journal} {\bibinfo  {journal} {New Journal of Physics}\ }\textbf
  {\bibinfo {volume} {13}},\ \bibinfo {pages} {053016} (\bibinfo {year}
  {2011})}\BibitemShut {NoStop}%
\bibitem [{\citenamefont {Fidkowski}\ \emph {et~al.}(2012)\citenamefont
  {Fidkowski}, \citenamefont {Alicea}, \citenamefont {Lindner}, \citenamefont
  {Lutchyn},\ and\ \citenamefont {Fisher}}]{Fidkowski2012}%
  \BibitemOpen
  \bibfield  {author} {\bibinfo {author} {\bibfnamefont {L.}~\bibnamefont
  {Fidkowski}}, \bibinfo {author} {\bibfnamefont {J.}~\bibnamefont {Alicea}},
  \bibinfo {author} {\bibfnamefont {N.~H.}\ \bibnamefont {Lindner}}, \bibinfo
  {author} {\bibfnamefont {R.~M.}\ \bibnamefont {Lutchyn}}, \ and\ \bibinfo
  {author} {\bibfnamefont {M.~P.~A.}\ \bibnamefont {Fisher}},\ }\href {\doibase
  10.1103/PhysRevB.85.245121} {\bibfield  {journal} {\bibinfo  {journal} {Phys.
  Rev. B}\ }\textbf {\bibinfo {volume} {85}},\ \bibinfo {pages} {245121}
  (\bibinfo {year} {2012})}\BibitemShut {NoStop}%
\bibitem [{\citenamefont {Mourik}\ \emph {et~al.}(2012)\citenamefont {Mourik},
  \citenamefont {Zuo}, \citenamefont {Frolov}, \citenamefont {Plissard},
  \citenamefont {Bakkers},\ and\ \citenamefont {Kouwenhoven}}]{Mourik2012}%
  \BibitemOpen
  \bibfield  {author} {\bibinfo {author} {\bibfnamefont {V.}~\bibnamefont
  {Mourik}}, \bibinfo {author} {\bibfnamefont {K.}~\bibnamefont {Zuo}},
  \bibinfo {author} {\bibfnamefont {S.~M.}\ \bibnamefont {Frolov}}, \bibinfo
  {author} {\bibfnamefont {S.~R.}\ \bibnamefont {Plissard}}, \bibinfo {author}
  {\bibfnamefont {E.~P. A.~M.}\ \bibnamefont {Bakkers}}, \ and\ \bibinfo
  {author} {\bibfnamefont {L.~P.}\ \bibnamefont {Kouwenhoven}},\ }\href
  {\doibase 10.1126/science.1222360} {\bibfield  {journal} {\bibinfo  {journal}
  {Science}\ }\textbf {\bibinfo {volume} {336}},\ \bibinfo {pages} {1003}
  (\bibinfo {year} {2012})}\BibitemShut {NoStop}%
\bibitem [{\citenamefont {Deng}\ \emph {et~al.}(2012)\citenamefont {Deng},
  \citenamefont {Yu}, \citenamefont {Huang}, \citenamefont {Larsson},
  \citenamefont {Caroff},\ and\ \citenamefont {Xu}}]{Deng2012}%
  \BibitemOpen
  \bibfield  {author} {\bibinfo {author} {\bibfnamefont {M.~T.}\ \bibnamefont
  {Deng}}, \bibinfo {author} {\bibfnamefont {C.~L.}\ \bibnamefont {Yu}},
  \bibinfo {author} {\bibfnamefont {G.~Y.}\ \bibnamefont {Huang}}, \bibinfo
  {author} {\bibfnamefont {M.}~\bibnamefont {Larsson}}, \bibinfo {author}
  {\bibfnamefont {P.}~\bibnamefont {Caroff}}, \ and\ \bibinfo {author}
  {\bibfnamefont {H.~Q.}\ \bibnamefont {Xu}},\ }\href {\doibase
  10.1021/nl303758w} {\bibfield  {journal} {\bibinfo  {journal} {Nano Letters}\
  }\textbf {\bibinfo {volume} {12}},\ \bibinfo {pages} {6414} (\bibinfo {year}
  {2012})}\BibitemShut {NoStop}%
\bibitem [{\citenamefont {Das}\ \emph {et~al.}(2012)\citenamefont {Das},
  \citenamefont {Ronen}, \citenamefont {Most}, \citenamefont {Oreg},
  \citenamefont {Heiblum},\ and\ \citenamefont {Shtrikman}}]{Das2012}%
  \BibitemOpen
  \bibfield  {author} {\bibinfo {author} {\bibfnamefont {A.}~\bibnamefont
  {Das}}, \bibinfo {author} {\bibfnamefont {Y.}~\bibnamefont {Ronen}}, \bibinfo
  {author} {\bibfnamefont {Y.}~\bibnamefont {Most}}, \bibinfo {author}
  {\bibfnamefont {Y.}~\bibnamefont {Oreg}}, \bibinfo {author} {\bibfnamefont
  {M.}~\bibnamefont {Heiblum}}, \ and\ \bibinfo {author} {\bibfnamefont
  {H.}~\bibnamefont {Shtrikman}},\ }\href@noop {} {\bibfield  {journal}
  {\bibinfo  {journal} {Nature Physics}\ }\textbf {\bibinfo {volume} {8}},\
  \bibinfo {pages} {887} (\bibinfo {year} {2012})}\BibitemShut {NoStop}%
\bibitem [{\citenamefont {Finck}\ \emph {et~al.}(2013)\citenamefont {Finck},
  \citenamefont {Van~Harlingen}, \citenamefont {Mohseni}, \citenamefont
  {Jung},\ and\ \citenamefont {Li}}]{Finck2013}%
  \BibitemOpen
  \bibfield  {author} {\bibinfo {author} {\bibfnamefont {A.~D.~K.}\
  \bibnamefont {Finck}}, \bibinfo {author} {\bibfnamefont {D.~J.}\ \bibnamefont
  {Van~Harlingen}}, \bibinfo {author} {\bibfnamefont {P.~K.}\ \bibnamefont
  {Mohseni}}, \bibinfo {author} {\bibfnamefont {K.}~\bibnamefont {Jung}}, \
  and\ \bibinfo {author} {\bibfnamefont {X.}~\bibnamefont {Li}},\ }\href
  {\doibase 10.1103/PhysRevLett.110.126406} {\bibfield  {journal} {\bibinfo
  {journal} {Phys. Rev. Lett.}\ }\textbf {\bibinfo {volume} {110}},\ \bibinfo
  {pages} {126406} (\bibinfo {year} {2013})}\BibitemShut {NoStop}%
\bibitem [{\citenamefont {Chang}\ \emph {et~al.}(2015)\citenamefont {Chang},
  \citenamefont {Albrecht}, \citenamefont {Jespersen}, \citenamefont
  {Kuemmeth}, \citenamefont {Krogstrup}, \citenamefont {Nyg{\r a}rd},\ and\
  \citenamefont {Marcus}}]{Chang2015}%
  \BibitemOpen
  \bibfield  {author} {\bibinfo {author} {\bibfnamefont {W.}~\bibnamefont
  {Chang}}, \bibinfo {author} {\bibfnamefont {S.~M.}\ \bibnamefont {Albrecht}},
  \bibinfo {author} {\bibfnamefont {T.~S.}\ \bibnamefont {Jespersen}}, \bibinfo
  {author} {\bibfnamefont {F.}~\bibnamefont {Kuemmeth}}, \bibinfo {author}
  {\bibfnamefont {P.}~\bibnamefont {Krogstrup}}, \bibinfo {author}
  {\bibfnamefont {J.}~\bibnamefont {Nyg{\r a}rd}}, \ and\ \bibinfo {author}
  {\bibfnamefont {C.~M.}\ \bibnamefont {Marcus}},\ }\href {\doibase
  10.1038/nnano.2014.306} {\bibfield  {journal} {\bibinfo  {journal} {Nat
  Nano}\ }\textbf {\bibinfo {volume} {10}},\ \bibinfo {pages} {232} (\bibinfo
  {year} {2015})}\BibitemShut {NoStop}%
\bibitem [{\citenamefont {Albrecht}\ \emph {et~al.}(2016)\citenamefont
  {Albrecht}, \citenamefont {Higginbotham}, \citenamefont {Madsen},
  \citenamefont {Kuemmeth}, \citenamefont {Jespersen}, \citenamefont {Nyg{\r
  a}rd}, \citenamefont {Krogstrup},\ and\ \citenamefont
  {Marcus}}]{Albrecht2016}%
  \BibitemOpen
  \bibfield  {author} {\bibinfo {author} {\bibfnamefont {S.~M.}\ \bibnamefont
  {Albrecht}}, \bibinfo {author} {\bibfnamefont {A.~P.}\ \bibnamefont
  {Higginbotham}}, \bibinfo {author} {\bibfnamefont {M.}~\bibnamefont
  {Madsen}}, \bibinfo {author} {\bibfnamefont {F.}~\bibnamefont {Kuemmeth}},
  \bibinfo {author} {\bibfnamefont {T.~S.}\ \bibnamefont {Jespersen}}, \bibinfo
  {author} {\bibfnamefont {J.}~\bibnamefont {Nyg{\r a}rd}}, \bibinfo {author}
  {\bibfnamefont {P.}~\bibnamefont {Krogstrup}}, \ and\ \bibinfo {author}
  {\bibfnamefont {C.~M.}\ \bibnamefont {Marcus}},\ }\href {\doibase
  10.1038/nature17162} {\bibfield  {journal} {\bibinfo  {journal} {Nature}\
  }\textbf {\bibinfo {volume} {531}},\ \bibinfo {pages} {206} (\bibinfo {year}
  {2016})}\BibitemShut {NoStop}%
\bibitem [{\citenamefont {Zhang}\ \emph {et~al.}(2016)\citenamefont {Zhang},
  \citenamefont {Gul}, \citenamefont {Conesa-Boj}, \citenamefont {Zuo},
  \citenamefont {Mourik}, \citenamefont {de~Vries}, \citenamefont {van Veen},
  \citenamefont {van Woerkom}, \citenamefont {Nowak}, \citenamefont {Wimmer},
  \citenamefont {Car}, \citenamefont {Plissard}, \citenamefont {Bakkers},
  \citenamefont {Quintero-Pérez}, \citenamefont {Goswami}, \citenamefont
  {Watanabe}, \citenamefont {Taniguchi},\ and\ \citenamefont
  {Kouwenhoven}}]{Zhang2016}%
  \BibitemOpen
  \bibfield  {author} {\bibinfo {author} {\bibfnamefont {H.}~\bibnamefont
  {Zhang}}, \bibinfo {author} {\bibfnamefont {O.}~\bibnamefont {Gul}}, \bibinfo
  {author} {\bibfnamefont {S.}~\bibnamefont {Conesa-Boj}}, \bibinfo {author}
  {\bibfnamefont {K.}~\bibnamefont {Zuo}}, \bibinfo {author} {\bibfnamefont
  {V.}~\bibnamefont {Mourik}}, \bibinfo {author} {\bibfnamefont {F.~K.}\
  \bibnamefont {de~Vries}}, \bibinfo {author} {\bibfnamefont {J.}~\bibnamefont
  {van Veen}}, \bibinfo {author} {\bibfnamefont {D.~J.}\ \bibnamefont {van
  Woerkom}}, \bibinfo {author} {\bibfnamefont {M.~P.}\ \bibnamefont {Nowak}},
  \bibinfo {author} {\bibfnamefont {M.}~\bibnamefont {Wimmer}}, \bibinfo
  {author} {\bibfnamefont {D.}~\bibnamefont {Car}}, \bibinfo {author}
  {\bibfnamefont {S.}~\bibnamefont {Plissard}}, \bibinfo {author}
  {\bibfnamefont {E.~P. A.~M.}\ \bibnamefont {Bakkers}}, \bibinfo {author}
  {\bibfnamefont {M.}~\bibnamefont {Quintero-Pérez}}, \bibinfo {author}
  {\bibfnamefont {S.}~\bibnamefont {Goswami}}, \bibinfo {author} {\bibfnamefont
  {K.}~\bibnamefont {Watanabe}}, \bibinfo {author} {\bibfnamefont
  {T.}~\bibnamefont {Taniguchi}}, \ and\ \bibinfo {author} {\bibfnamefont
  {L.~P.}\ \bibnamefont {Kouwenhoven}},\ }\href
  {https://arxiv.org/abs/1603.04069} {\bibfield  {journal} {\bibinfo  {journal}
  {e-print arXiv:1603.04069}\ } (\bibinfo {year} {2016})}\BibitemShut {NoStop}%
\bibitem [{\citenamefont {Deng}\ \emph {et~al.}(2016)\citenamefont {Deng},
  \citenamefont {Vaitiekenas}, \citenamefont {Hansen}, \citenamefont {Danon},
  \citenamefont {Leijnse}, \citenamefont {Flensberg}, \citenamefont {Nyg{\r
  a}rd}, \citenamefont {Krogstrup},\ and\ \citenamefont {Marcus}}]{Deng2016}%
  \BibitemOpen
  \bibfield  {author} {\bibinfo {author} {\bibfnamefont {M.~T.}\ \bibnamefont
  {Deng}}, \bibinfo {author} {\bibfnamefont {S.}~\bibnamefont {Vaitiekenas}},
  \bibinfo {author} {\bibfnamefont {E.~B.}\ \bibnamefont {Hansen}}, \bibinfo
  {author} {\bibfnamefont {J.}~\bibnamefont {Danon}}, \bibinfo {author}
  {\bibfnamefont {M.}~\bibnamefont {Leijnse}}, \bibinfo {author} {\bibfnamefont
  {K.}~\bibnamefont {Flensberg}}, \bibinfo {author} {\bibfnamefont
  {J.}~\bibnamefont {Nyg{\r a}rd}}, \bibinfo {author} {\bibfnamefont
  {P.}~\bibnamefont {Krogstrup}}, \ and\ \bibinfo {author} {\bibfnamefont
  {C.~M.}\ \bibnamefont {Marcus}},\ }\href {\doibase 10.1126/science.aaf3961}
  {\bibfield  {journal} {\bibinfo  {journal} {Science}\ }\textbf {\bibinfo
  {volume} {354}},\ \bibinfo {pages} {1557} (\bibinfo {year}
  {2016})}\BibitemShut {NoStop}%
\bibitem [{\citenamefont {Chen}\ \emph {et~al.}(2016)\citenamefont {Chen},
  \citenamefont {Yu}, \citenamefont {Stenger}, \citenamefont {Hocevar},
  \citenamefont {Car}, \citenamefont {Plissard}, \citenamefont {Bakkers},
  \citenamefont {Stanescu},\ and\ \citenamefont {Frolov}}]{Chen2016}%
  \BibitemOpen
  \bibfield  {author} {\bibinfo {author} {\bibfnamefont {J.}~\bibnamefont
  {Chen}}, \bibinfo {author} {\bibfnamefont {P.}~\bibnamefont {Yu}}, \bibinfo
  {author} {\bibfnamefont {J.}~\bibnamefont {Stenger}}, \bibinfo {author}
  {\bibfnamefont {M.}~\bibnamefont {Hocevar}}, \bibinfo {author} {\bibfnamefont
  {D.}~\bibnamefont {Car}}, \bibinfo {author} {\bibfnamefont {S.~R.}\
  \bibnamefont {Plissard}}, \bibinfo {author} {\bibfnamefont {E.~P.}\
  \bibnamefont {Bakkers}}, \bibinfo {author} {\bibfnamefont {T.~D.}\
  \bibnamefont {Stanescu}}, \ and\ \bibinfo {author} {\bibfnamefont {S.~M.}\
  \bibnamefont {Frolov}},\ }\href {https://arxiv.org/abs/1610.04555} {\bibfield
   {journal} {\bibinfo  {journal} {e-print arXiv:1610.04555}\ } (\bibinfo
  {year} {2016})}\BibitemShut {NoStop}%
\bibitem [{\citenamefont {Stanescu}\ \emph {et~al.}(2011)\citenamefont
  {Stanescu}, \citenamefont {Lutchyn},\ and\ \citenamefont
  {Das~Sarma}}]{Stanescu2011}%
  \BibitemOpen
  \bibfield  {author} {\bibinfo {author} {\bibfnamefont {T.~D.}\ \bibnamefont
  {Stanescu}}, \bibinfo {author} {\bibfnamefont {R.~M.}\ \bibnamefont
  {Lutchyn}}, \ and\ \bibinfo {author} {\bibfnamefont {S.}~\bibnamefont
  {Das~Sarma}},\ }\href {\doibase 10.1103/PhysRevB.84.144522} {\bibfield
  {journal} {\bibinfo  {journal} {Phys. Rev. B}\ }\textbf {\bibinfo {volume}
  {84}},\ \bibinfo {pages} {144522} (\bibinfo {year} {2011})}\BibitemShut
  {NoStop}%
\bibitem [{\citenamefont {Pientka}\ \emph {et~al.}(2012)\citenamefont
  {Pientka}, \citenamefont {Kells}, \citenamefont {Romito}, \citenamefont
  {Brouwer},\ and\ \citenamefont {von Oppen}}]{Pientka2012}%
  \BibitemOpen
  \bibfield  {author} {\bibinfo {author} {\bibfnamefont {F.}~\bibnamefont
  {Pientka}}, \bibinfo {author} {\bibfnamefont {G.}~\bibnamefont {Kells}},
  \bibinfo {author} {\bibfnamefont {A.}~\bibnamefont {Romito}}, \bibinfo
  {author} {\bibfnamefont {P.~W.}\ \bibnamefont {Brouwer}}, \ and\ \bibinfo
  {author} {\bibfnamefont {F.}~\bibnamefont {von Oppen}},\ }\href {\doibase
  10.1103/PhysRevLett.109.227006} {\bibfield  {journal} {\bibinfo  {journal}
  {Phys. Rev. Lett.}\ }\textbf {\bibinfo {volume} {109}},\ \bibinfo {pages}
  {227006} (\bibinfo {year} {2012})}\BibitemShut {NoStop}%
\bibitem [{\citenamefont {Prada}\ \emph {et~al.}(2012)\citenamefont {Prada},
  \citenamefont {San-Jose},\ and\ \citenamefont {Aguado}}]{Prada2012}%
  \BibitemOpen
  \bibfield  {author} {\bibinfo {author} {\bibfnamefont {E.}~\bibnamefont
  {Prada}}, \bibinfo {author} {\bibfnamefont {P.}~\bibnamefont {San-Jose}}, \
  and\ \bibinfo {author} {\bibfnamefont {R.}~\bibnamefont {Aguado}},\ }\href
  {\doibase 10.1103/PhysRevB.86.180503} {\bibfield  {journal} {\bibinfo
  {journal} {Phys. Rev. B}\ }\textbf {\bibinfo {volume} {86}},\ \bibinfo
  {pages} {180503} (\bibinfo {year} {2012})}\BibitemShut {NoStop}%
\bibitem [{\citenamefont {Lin}\ \emph {et~al.}(2012)\citenamefont {Lin},
  \citenamefont {Sau},\ and\ \citenamefont {Das~Sarma}}]{Lin2012}%
  \BibitemOpen
  \bibfield  {author} {\bibinfo {author} {\bibfnamefont {C.-H.}\ \bibnamefont
  {Lin}}, \bibinfo {author} {\bibfnamefont {J.~D.}\ \bibnamefont {Sau}}, \ and\
  \bibinfo {author} {\bibfnamefont {S.}~\bibnamefont {Das~Sarma}},\ }\href
  {\doibase 10.1103/PhysRevB.86.224511} {\bibfield  {journal} {\bibinfo
  {journal} {Phys. Rev. B}\ }\textbf {\bibinfo {volume} {86}},\ \bibinfo
  {pages} {224511} (\bibinfo {year} {2012})}\BibitemShut {NoStop}%
\bibitem [{\citenamefont {Rainis}\ \emph {et~al.}(2013)\citenamefont {Rainis},
  \citenamefont {Trifunovic}, \citenamefont {Klinovaja},\ and\ \citenamefont
  {Loss}}]{Rainis2013}%
  \BibitemOpen
  \bibfield  {author} {\bibinfo {author} {\bibfnamefont {D.}~\bibnamefont
  {Rainis}}, \bibinfo {author} {\bibfnamefont {L.}~\bibnamefont {Trifunovic}},
  \bibinfo {author} {\bibfnamefont {J.}~\bibnamefont {Klinovaja}}, \ and\
  \bibinfo {author} {\bibfnamefont {D.}~\bibnamefont {Loss}},\ }\href {\doibase
  10.1103/PhysRevB.87.024515} {\bibfield  {journal} {\bibinfo  {journal} {Phys.
  Rev. B}\ }\textbf {\bibinfo {volume} {87}},\ \bibinfo {pages} {024515}
  (\bibinfo {year} {2013})}\BibitemShut {NoStop}%
\bibitem [{\citenamefont {Stanescu}\ \emph {et~al.}(2014)\citenamefont
  {Stanescu}, \citenamefont {Lutchyn},\ and\ \citenamefont
  {Das~Sarma}}]{Stanescu2014}%
  \BibitemOpen
  \bibfield  {author} {\bibinfo {author} {\bibfnamefont {T.~D.}\ \bibnamefont
  {Stanescu}}, \bibinfo {author} {\bibfnamefont {R.~M.}\ \bibnamefont
  {Lutchyn}}, \ and\ \bibinfo {author} {\bibfnamefont {S.}~\bibnamefont
  {Das~Sarma}},\ }\href {\doibase 10.1103/PhysRevB.90.085302} {\bibfield
  {journal} {\bibinfo  {journal} {Phys. Rev. B}\ }\textbf {\bibinfo {volume}
  {90}},\ \bibinfo {pages} {085302} (\bibinfo {year} {2014})}\BibitemShut
  {NoStop}%
\bibitem [{\citenamefont {Yan}\ and\ \citenamefont {Wan}(2014)}]{Yan2014}%
  \BibitemOpen
  \bibfield  {author} {\bibinfo {author} {\bibfnamefont {Z.}~\bibnamefont
  {Yan}}\ and\ \bibinfo {author} {\bibfnamefont {S.}~\bibnamefont {Wan}},\
  }\href {http://stacks.iop.org/1367-2630/16/i=9/a=093004} {\bibfield
  {journal} {\bibinfo  {journal} {New Journal of Physics}\ }\textbf {\bibinfo
  {volume} {16}},\ \bibinfo {pages} {093004} (\bibinfo {year}
  {2014})}\BibitemShut {NoStop}%
\bibitem [{\citenamefont {Setiawan}\ \emph {et~al.}(2015)\citenamefont
  {Setiawan}, \citenamefont {Brydon}, \citenamefont {Sau},\ and\ \citenamefont
  {Das~Sarma}}]{Setiwan2015}%
  \BibitemOpen
  \bibfield  {author} {\bibinfo {author} {\bibfnamefont {F.}~\bibnamefont
  {Setiawan}}, \bibinfo {author} {\bibfnamefont {P.~M.~R.}\ \bibnamefont
  {Brydon}}, \bibinfo {author} {\bibfnamefont {J.~D.}\ \bibnamefont {Sau}}, \
  and\ \bibinfo {author} {\bibfnamefont {S.}~\bibnamefont {Das~Sarma}},\ }\href
  {\doibase 10.1103/PhysRevB.91.214513} {\bibfield  {journal} {\bibinfo
  {journal} {Phys. Rev. B}\ }\textbf {\bibinfo {volume} {91}},\ \bibinfo
  {pages} {214513} (\bibinfo {year} {2015})}\BibitemShut {NoStop}%
\bibitem [{\citenamefont {Liu}\ \emph {et~al.}(2017)\citenamefont {Liu},
  \citenamefont {Sau},\ and\ \citenamefont {Das~Sarma}}]{Liu2017}%
  \BibitemOpen
  \bibfield  {author} {\bibinfo {author} {\bibfnamefont {C.-X.}\ \bibnamefont
  {Liu}}, \bibinfo {author} {\bibfnamefont {J.~D.}\ \bibnamefont {Sau}}, \ and\
  \bibinfo {author} {\bibfnamefont {S.}~\bibnamefont {Das~Sarma}},\ }\href
  {\doibase 10.1103/PhysRevB.95.054502} {\bibfield  {journal} {\bibinfo
  {journal} {Phys. Rev. B}\ }\textbf {\bibinfo {volume} {95}},\ \bibinfo
  {pages} {054502} (\bibinfo {year} {2017})}\BibitemShut {NoStop}%
\bibitem [{\citenamefont {Peng}\ \emph {et~al.}(2015)\citenamefont {Peng},
  \citenamefont {Pientka}, \citenamefont {Vinkler-Aviv}, \citenamefont
  {Glazman},\ and\ \citenamefont {von Oppen}}]{Peng2015}%
  \BibitemOpen
  \bibfield  {author} {\bibinfo {author} {\bibfnamefont {Y.}~\bibnamefont
  {Peng}}, \bibinfo {author} {\bibfnamefont {F.}~\bibnamefont {Pientka}},
  \bibinfo {author} {\bibfnamefont {Y.}~\bibnamefont {Vinkler-Aviv}}, \bibinfo
  {author} {\bibfnamefont {L.~I.}\ \bibnamefont {Glazman}}, \ and\ \bibinfo
  {author} {\bibfnamefont {F.}~\bibnamefont {von Oppen}},\ }\href {\doibase
  10.1103/PhysRevLett.115.266804} {\bibfield  {journal} {\bibinfo  {journal}
  {Phys. Rev. Lett.}\ }\textbf {\bibinfo {volume} {115}},\ \bibinfo {pages}
  {266804} (\bibinfo {year} {2015})}\BibitemShut {NoStop}%
\bibitem [{\citenamefont {Sharma}\ and\ \citenamefont
  {Tewari}(2016)}]{Sharma2016}%
  \BibitemOpen
  \bibfield  {author} {\bibinfo {author} {\bibfnamefont {G.}~\bibnamefont
  {Sharma}}\ and\ \bibinfo {author} {\bibfnamefont {S.}~\bibnamefont
  {Tewari}},\ }\href {\doibase 10.1103/PhysRevB.93.195161} {\bibfield
  {journal} {\bibinfo  {journal} {Phys. Rev. B}\ }\textbf {\bibinfo {volume}
  {93}},\ \bibinfo {pages} {195161} (\bibinfo {year} {2016})}\BibitemShut
  {NoStop}%
\bibitem [{\citenamefont {Chevallier}\ and\ \citenamefont
  {Klinovaja}(2016)}]{Chevallier2016}%
  \BibitemOpen
  \bibfield  {author} {\bibinfo {author} {\bibfnamefont {D.}~\bibnamefont
  {Chevallier}}\ and\ \bibinfo {author} {\bibfnamefont {J.}~\bibnamefont
  {Klinovaja}},\ }\href {\doibase 10.1103/PhysRevB.94.035417} {\bibfield
  {journal} {\bibinfo  {journal} {Phys. Rev. B}\ }\textbf {\bibinfo {volume}
  {94}},\ \bibinfo {pages} {035417} (\bibinfo {year} {2016})}\BibitemShut
  {NoStop}%
\bibitem [{\citenamefont {Setiawan}\ \emph {et~al.}(2017)\citenamefont
  {Setiawan}, \citenamefont {Cole}, \citenamefont {Sau},\ and\ \citenamefont
  {Das~Sarma}}]{Setiawan2017}%
  \BibitemOpen
  \bibfield  {author} {\bibinfo {author} {\bibfnamefont {F.}~\bibnamefont
  {Setiawan}}, \bibinfo {author} {\bibfnamefont {W.~S.}\ \bibnamefont {Cole}},
  \bibinfo {author} {\bibfnamefont {J.~D.}\ \bibnamefont {Sau}}, \ and\
  \bibinfo {author} {\bibfnamefont {S.}~\bibnamefont {Das~Sarma}},\ }\href
  {\doibase 10.1103/PhysRevB.95.020501} {\bibfield  {journal} {\bibinfo
  {journal} {Phys. Rev. B}\ }\textbf {\bibinfo {volume} {95}},\ \bibinfo
  {pages} {020501} (\bibinfo {year} {2017})}\BibitemShut {NoStop}%
\bibitem [{\citenamefont {Reeg}\ and\ \citenamefont {Maslov}(2017)}]{Reeg2017}%
  \BibitemOpen
  \bibfield  {author} {\bibinfo {author} {\bibfnamefont {C.}~\bibnamefont
  {Reeg}}\ and\ \bibinfo {author} {\bibfnamefont {D.~L.}\ \bibnamefont
  {Maslov}},\ }\href {\doibase 10.1103/PhysRevB.95.205439} {\bibfield
  {journal} {\bibinfo  {journal} {Phys. Rev. B}\ }\textbf {\bibinfo {volume}
  {95}},\ \bibinfo {pages} {205439} (\bibinfo {year} {2017})}\BibitemShut
  {NoStop}%
\bibitem [{\citenamefont {Stanescu}\ and\ \citenamefont
  {Sarma}(2017)}]{Stanescu2017a}%
  \BibitemOpen
  \bibfield  {author} {\bibinfo {author} {\bibfnamefont {T.~D.}\ \bibnamefont
  {Stanescu}}\ and\ \bibinfo {author} {\bibfnamefont {S.~D.}\ \bibnamefont
  {Sarma}},\ }\href {https://arxiv.org/abs/1702.03976} {\bibfield  {journal}
  {\bibinfo  {journal} {e-print arXiv:1702.03976}\ } (\bibinfo {year}
  {2017})}\BibitemShut {NoStop}%
\bibitem [{\citenamefont {Blonder}\ \emph {et~al.}(1982)\citenamefont
  {Blonder}, \citenamefont {Tinkham},\ and\ \citenamefont
  {Klapwijk}}]{Blonder1982}%
  \BibitemOpen
  \bibfield  {author} {\bibinfo {author} {\bibfnamefont {G.~E.}\ \bibnamefont
  {Blonder}}, \bibinfo {author} {\bibfnamefont {M.}~\bibnamefont {Tinkham}}, \
  and\ \bibinfo {author} {\bibfnamefont {T.~M.}\ \bibnamefont {Klapwijk}},\
  }\href {\doibase 10.1103/PhysRevB.25.4515} {\bibfield  {journal} {\bibinfo
  {journal} {Phys. Rev. B}\ }\textbf {\bibinfo {volume} {25}},\ \bibinfo
  {pages} {4515} (\bibinfo {year} {1982})}\BibitemShut {NoStop}%
\bibitem [{\citenamefont {Rammer}\ and\ \citenamefont
  {Smith}(1986)}]{Rammer1986}%
  \BibitemOpen
  \bibfield  {author} {\bibinfo {author} {\bibfnamefont {J.}~\bibnamefont
  {Rammer}}\ and\ \bibinfo {author} {\bibfnamefont {H.}~\bibnamefont {Smith}},\
  }\href {\doibase 10.1103/RevModPhys.58.323} {\bibfield  {journal} {\bibinfo
  {journal} {Rev. Mod. Phys.}\ }\textbf {\bibinfo {volume} {58}},\ \bibinfo
  {pages} {323} (\bibinfo {year} {1986})}\BibitemShut {NoStop}%
\bibitem [{\citenamefont {Sticlet}\ \emph {et~al.}(2016)\citenamefont
  {Sticlet}, \citenamefont {Nijholt},\ and\ \citenamefont
  {Akhmerov}}]{Sticlet2016}%
  \BibitemOpen
  \bibfield  {author} {\bibinfo {author} {\bibfnamefont {D.}~\bibnamefont
  {Sticlet}}, \bibinfo {author} {\bibfnamefont {B.}~\bibnamefont {Nijholt}}, \
  and\ \bibinfo {author} {\bibfnamefont {A.}~\bibnamefont {Akhmerov}},\ }\href
  {https://arxiv.org/abs/1609.00637} {\bibfield  {journal} {\bibinfo  {journal}
  {e-print arXiv:1609.00637}\ } (\bibinfo {year} {2016})}\BibitemShut {NoStop}%
\bibitem [{\citenamefont {Berthod}\ and\ \citenamefont
  {Giamarchi}(2011)}]{Berthod2011}%
  \BibitemOpen
  \bibfield  {author} {\bibinfo {author} {\bibfnamefont {C.}~\bibnamefont
  {Berthod}}\ and\ \bibinfo {author} {\bibfnamefont {T.}~\bibnamefont
  {Giamarchi}},\ }\href {\doibase 10.1103/PhysRevB.84.155414} {\bibfield
  {journal} {\bibinfo  {journal} {Phys. Rev. B}\ }\textbf {\bibinfo {volume}
  {84}},\ \bibinfo {pages} {155414} (\bibinfo {year} {2011})}\BibitemShut
  {NoStop}%
\bibitem [{\citenamefont {He}\ \emph {et~al.}(2014)\citenamefont {He},
  \citenamefont {Ng}, \citenamefont {Lee},\ and\ \citenamefont {Law}}]{He2014}%
  \BibitemOpen
  \bibfield  {author} {\bibinfo {author} {\bibfnamefont {J.~J.}\ \bibnamefont
  {He}}, \bibinfo {author} {\bibfnamefont {T.~K.}\ \bibnamefont {Ng}}, \bibinfo
  {author} {\bibfnamefont {P.~A.}\ \bibnamefont {Lee}}, \ and\ \bibinfo
  {author} {\bibfnamefont {K.~T.}\ \bibnamefont {Law}},\ }\href {\doibase
  10.1103/PhysRevLett.112.037001} {\bibfield  {journal} {\bibinfo  {journal}
  {Phys. Rev. Lett.}\ }\textbf {\bibinfo {volume} {112}},\ \bibinfo {pages}
  {037001} (\bibinfo {year} {2014})}\BibitemShut {NoStop}%
\bibitem [{\citenamefont {Cole}\ \emph {et~al.}(2015)\citenamefont {Cole},
  \citenamefont {Das~Sarma},\ and\ \citenamefont {Stanescu}}]{Cole2015}%
  \BibitemOpen
  \bibfield  {author} {\bibinfo {author} {\bibfnamefont {W.~S.}\ \bibnamefont
  {Cole}}, \bibinfo {author} {\bibfnamefont {S.}~\bibnamefont {Das~Sarma}}, \
  and\ \bibinfo {author} {\bibfnamefont {T.~D.}\ \bibnamefont {Stanescu}},\
  }\href {\doibase 10.1103/PhysRevB.92.174511} {\bibfield  {journal} {\bibinfo
  {journal} {Phys. Rev. B}\ }\textbf {\bibinfo {volume} {92}},\ \bibinfo
  {pages} {174511} (\bibinfo {year} {2015})}\BibitemShut {NoStop}%
\bibitem [{\citenamefont {Cole}\ \emph {et~al.}(2016)\citenamefont {Cole},
  \citenamefont {Sau},\ and\ \citenamefont {Das~Sarma}}]{Cole2016}%
  \BibitemOpen
  \bibfield  {author} {\bibinfo {author} {\bibfnamefont {W.~S.}\ \bibnamefont
  {Cole}}, \bibinfo {author} {\bibfnamefont {J.~D.}\ \bibnamefont {Sau}}, \
  and\ \bibinfo {author} {\bibfnamefont {S.}~\bibnamefont {Das~Sarma}},\ }\href
  {\doibase 10.1103/PhysRevB.94.140505} {\bibfield  {journal} {\bibinfo
  {journal} {Phys. Rev. B}\ }\textbf {\bibinfo {volume} {94}},\ \bibinfo
  {pages} {140505} (\bibinfo {year} {2016})}\BibitemShut {NoStop}%
\bibitem [{\citenamefont {Hui}\ \emph {et~al.}(2015)\citenamefont {Hui},
  \citenamefont {Sau},\ and\ \citenamefont {Das~Sarma}}]{Hui2015}%
  \BibitemOpen
  \bibfield  {author} {\bibinfo {author} {\bibfnamefont {H.-Y.}\ \bibnamefont
  {Hui}}, \bibinfo {author} {\bibfnamefont {J.~D.}\ \bibnamefont {Sau}}, \ and\
  \bibinfo {author} {\bibfnamefont {S.}~\bibnamefont {Das~Sarma}},\ }\href
  {\doibase 10.1103/PhysRevB.92.174512} {\bibfield  {journal} {\bibinfo
  {journal} {Phys. Rev. B}\ }\textbf {\bibinfo {volume} {92}},\ \bibinfo
  {pages} {174512} (\bibinfo {year} {2015})}\BibitemShut {NoStop}%
\bibitem [{\citenamefont {Martin}\ and\ \citenamefont
  {Mozyrsky}(2014)}]{Martin2014}%
  \BibitemOpen
  \bibfield  {author} {\bibinfo {author} {\bibfnamefont {I.}~\bibnamefont
  {Martin}}\ and\ \bibinfo {author} {\bibfnamefont {D.}~\bibnamefont
  {Mozyrsky}},\ }\href {\doibase 10.1103/PhysRevB.90.100508} {\bibfield
  {journal} {\bibinfo  {journal} {Phys. Rev. B}\ }\textbf {\bibinfo {volume}
  {90}},\ \bibinfo {pages} {100508} (\bibinfo {year} {2014})}\BibitemShut
  {NoStop}%
\bibitem [{\citenamefont {Bauriedl}\ \emph {et~al.}(1981)\citenamefont
  {Bauriedl}, \citenamefont {Ziemann},\ and\ \citenamefont
  {Buckel}}]{Bauriedl1981}%
  \BibitemOpen
  \bibfield  {author} {\bibinfo {author} {\bibfnamefont {W.}~\bibnamefont
  {Bauriedl}}, \bibinfo {author} {\bibfnamefont {P.}~\bibnamefont {Ziemann}}, \
  and\ \bibinfo {author} {\bibfnamefont {W.}~\bibnamefont {Buckel}},\ }\href
  {\doibase 10.1103/PhysRevLett.47.1163} {\bibfield  {journal} {\bibinfo
  {journal} {Phys. Rev. Lett.}\ }\textbf {\bibinfo {volume} {47}},\ \bibinfo
  {pages} {1163} (\bibinfo {year} {1981})}\BibitemShut {NoStop}%
\bibitem [{\citenamefont {Yazdani}\ \emph {et~al.}(1997)\citenamefont
  {Yazdani}, \citenamefont {Jones}, \citenamefont {Lutz}, \citenamefont
  {Crommie},\ and\ \citenamefont {Eigler}}]{Yazdani1997}%
  \BibitemOpen
  \bibfield  {author} {\bibinfo {author} {\bibfnamefont {A.}~\bibnamefont
  {Yazdani}}, \bibinfo {author} {\bibfnamefont {B.~A.}\ \bibnamefont {Jones}},
  \bibinfo {author} {\bibfnamefont {C.~P.}\ \bibnamefont {Lutz}}, \bibinfo
  {author} {\bibfnamefont {M.~F.}\ \bibnamefont {Crommie}}, \ and\ \bibinfo
  {author} {\bibfnamefont {D.~M.}\ \bibnamefont {Eigler}},\ }\href {\doibase
  10.1126/science.275.5307.1767} {\bibfield  {journal} {\bibinfo  {journal}
  {Science}\ }\textbf {\bibinfo {volume} {275}},\ \bibinfo {pages} {1767}
  (\bibinfo {year} {1997})},\ \Eprint
  {http://arxiv.org/abs/http://science.sciencemag.org/content/275/5307/1767.full.pdf}
  {http://science.sciencemag.org/content/275/5307/1767.full.pdf} \BibitemShut
  {NoStop}%
\bibitem [{\citenamefont {Das~Sarma}\ \emph {et~al.}(2016)\citenamefont
  {Das~Sarma}, \citenamefont {Nag},\ and\ \citenamefont {Sau}}]{DSarma2016}%
  \BibitemOpen
  \bibfield  {author} {\bibinfo {author} {\bibfnamefont {S.}~\bibnamefont
  {Das~Sarma}}, \bibinfo {author} {\bibfnamefont {A.}~\bibnamefont {Nag}}, \
  and\ \bibinfo {author} {\bibfnamefont {J.~D.}\ \bibnamefont {Sau}},\ }\href
  {\doibase 10.1103/PhysRevB.94.035143} {\bibfield  {journal} {\bibinfo
  {journal} {Phys. Rev. B}\ }\textbf {\bibinfo {volume} {94}},\ \bibinfo
  {pages} {035143} (\bibinfo {year} {2016})}\BibitemShut {NoStop}%
\end{thebibliography}%

\end{document}